\documentclass[useAMS,usenatbib]{mn2e}

\usepackage{graphicx}
\usepackage{txfonts}
\usepackage[german]{babel}
\usepackage{latexsym}

\title[On the interaction of a thin, supersonic shell with a molecular cloud]{On the interaction of a thin, supersonic shell with a molecular cloud}
\author[Anathpindika. S. and Bhatt, H. C.]{Anathpindika. S. $^{1,2}$\thanks{E-mail:
spxsva@astro.cf.ac.uk (SVA); hcbhatt@iiap.res.in (HCB)} and Bhatt. H. C$^{1}$\\
$^{1}$Indian Institute of Astrophysics, Bangalore 560034, India\\
$^{2}$School of Astronomy \& Physics, Cardiff university, 5-The Parade, CF24 3AA, Cardiff, UK}
\begin{document}

\date{Accepted 0000 December 00. Received 0000 December 00; in original form 0000 October 00}

\pagerange{\pageref{firstpage}--\pageref{lastpage}} \pubyear{2002}

\maketitle

\label{firstpage}

\begin{abstract}

Molecular clouds (MCs) are stellar nurseries, however, formation of stars within MCs depends on the ambient physical conditions. MCs, over a free-fall time are exposed to numerous dynamical phenomena, of which, the interaction with a thin, dense shell of gas is but one. Below we present results from self-gravitating, 3-D smoothed particle hydrodynamics ({\small SPH}) simulations of the problem; seven realisations of the problem have been performed by varying the precollision density within the cloud, the nature of the post-collision shock, and the spatial resolution in the computational domain. Irrespective of the type of shock, a complex network of dense filaments, seeded by numerical noise, readily appears in the shocked cloud. Segregation of the dense and rarefied gas phases also manifests itself in a bimodal distribution of gas density. We demonstrate that the power-spectrum for rarefied gas is Kolomogorov like, while that for the denser gas is considerably steeper. As a corollary to the main problem, we also look into the possibly degenerative effect of the {\small SPH} artificial viscosity on the impact of the incident shell. It is observed that stronger viscosity leads to greater post-shock dissipation, that strongly decelerates the incident shock-front and promotes formation of contiguous structure, albeit on a much longer timescale. We conclude that too much viscosity is likely to enhance the proclivity towards gravitational boundedness of structure, leading to unphysical fragmentation.On the other hand, insufficient resolution appears to suppress fragmentation. Convergence of results is tested at both extremes, first by repeating the test case with more than a million particles and then with only half the number of particles in the original test case.

\end{abstract}


\section{Introduction}

\emph{Expanding shells and triggered star-formation.} The stability of the galactic disk and its evolution depends on the availability of molecular gas, that can be converted in to stars. Thus, an important prelude to understanding galactic evolution is the determination of factors that could possibly control the rate of star-formation. The problem in turn behoves a study of the stellar feed-back mechanism and its effect on the efficiency of star-formation. On this latter issue there has been considerable harrumph leading to antipodal suggestions : one such view propounded by \citet{b16} for instance is that, stellar feed-back, and turbulence in general act to vitiate star-formation as a consequence of which the process may be retarded, resulting in rather inefficient star-formation. This is reflected in a small value of the star-formation efficiency calculated over a free-fall time (SFE$_{ff}$). This proposition has been contested by \citet{b17}, who suggest that turbulent motion assists rapid assembly of gas into dense clumps, some of which could perhaps spawn stars. This dynamical process of forming stars apparently occurs on a short time-scale, of about a crossing-time; a similar idea was previously proposed by \citet{b18}.\\ \\
\emph{Turbulence in the interstellar medium (ISM).} Large scale turbulence in the galactic disk, and within molecular clouds (MCs), appears to be self-similar and scale-free. Various physical properties such as the length-scale, and velocity dispersion of the observed turbulent motion were related by empirical expressions first suggested by \citet{b3}. Turbulent motion in the ISM, inferred via broadening of spectral emission lines, and believed to be crucial in producing local density enhancements, is often thought to be shock-induced. Fluctuations in the density field produces hierarchical structure, evident from the preponderance of dense filaments and clumps in the ISM. Some authors have previously also employed statistical techniques to study the density field, and in this connection, a particularly useful quantity is the probability distribution function (PDF) of gas density. The non-uniform nature of a turbulent density field for an isothermal gas has been demonstrated to be approximately lognormal with the aid of numerical simulations (e.g. V{\`a}zquez-Semadeni 1994; Padoan, Nordlund \& Jones 1997a; Scalo \emph{et al.} 1998; V{\`a}zquez-Semadeni \emph{et al.} 2008).  

In fact, some authors, using the the lognormal density PDF as a basis, have derived a general dense core mass function (CMF) and the stellar initial mass function (IMF), that is roughly lognormal with a slope consistent to that of the canonical IMF (e.g. Padoan 1995, Padoan, Nordlund \& Jones 1997b; Padoan \& Nordlund 2002). The latter authors in support of their hypothesis, have also shown that the density distribution in obscure dark-clouds reported by \citet{b11} was consistent with a lognormal density PDF. The apparent similarity in the nature of the density PDF with the stellar IMF, and/or the CMF, though intriguing, may not necessarily have a causal connection as has been pointed out by \citet{b8}. Yet others have sought to explain the origin of hierarchical structure through the influence of turbulence over fractal MCs (e.g. Elmegreen 1997; Elmegreen 2002). Turbulence in the ISM therefore plays a crucial part in creating density structures, and perchance, regulates the rate of star-formation in MCs; shocks supposedly responsible for generating turbulent velocity fields could be driven either by stellar winds or energetic jets, expanding dense gas shells or collision between energetic gas streams (e.g. see reviews by Elmegreen \& Scalo 2004; Mac Low \& Klessen 2004, and references therein). 

\emph{Expanding shells} An approximately spherical shell of powerful ionising radiation (the Stromgr{"en} sphere), driven by either a young star-cluster, a massive O star or a supernova, sweeps up gas in the ISM whence it acquires mass \citep{b19}. The impact of such a shell, moving supersonically relative to the ISM, with a MC could compress the latter, and possibly enhance star-formation in case the cloud is self-gravitating, or trigger star-formation in a quiescent MC \citep{b23,b26,b25}. The MC could even be disrupted and sheared apart in case the shock is too strong \citep{b24}. Although direct observations of expanding shells interacting with MCs are relatively scarce, there are a few cases where the region of impact has been well studied. \citet{b30} and other authors cited in that work, using emission maps due to shock excited CO molecules in the region M17 (the Omega Nebula), have suggested sequential star-formation to have been triggered due to the interaction of an HII shell with MCs in the region. Similarly, \citet{b20,b21} have studied the interaction between a MC and an SNR, W51C using the H I 21 cm line and other molecular line emissions. The post-collision shock produces filamentary structure in the gas, however, there is no evidence for any star-formation activity. The Vela SNR, shown in Fig. 1, is another well studied example, see for instance \citet{b22} and references there-in. The apparently filamentary and other spatially extensive bright spots visible in Fig. 1 are supposedly due to MCs engulfed by the SNR. A survey of shocked CO emission in the region by Moriguchi \emph{et al.} (2001) has shown that the average size of a MC in the Vela SNR region is about 1 pc, having mass of about 60 M$_{\odot}$; the smallest cloud is about 3 M$_{\odot}$ massive, with a size of about 0.3 pc. Recently \citet{b27} have reported star-formation activity in clouds S175, believed to be triggered by associated HII regions. 

In one of the first numerical studies of the problem, \citet{b36} using grid based calculation showed that a supersonic shock has a strong shearing interaction with the cloud, that is eventually ruptured by the resulting hydrodynamic instabilities. Similar results were reported by \citet{b37} and \citet{b38}, in case of magnetised clouds. \citet{b28} simulated the problem using self-gravitating smoothed particle hydrodynamics ({\small SPH}) and showed that self-gravity of the post-collision shocked gas was crucial for star-formation to commence in it; the MC is likely to diffuse away in case gravity remained subservient to thermal energy, in other words when the post-shock radiative cooling was inefficient. In the present work we discuss high resolution {\small SPH} simulations and wish to emphasise on the density distribution within the shocked cloud. While the previous work employing grid codes has shown that a strong shock flattens the post-collision cloud on a rather short timescale, without any other dynamical signature within the cloud, this work suggests, such is not the case. We argue that shock induced turbulence leads to structure formation within the shocked cloud, some of which shows signs of boundedness. Convergence of the results is tested by repeating one of the test cases with eight times the number of particles within the precollision cloud, and then with half the number of particles. We have, however, not tested the extreme case where the MC may be flattened by the impact of the incident shell.

The plan of the paper is as follows. We discuss the initial conditions in \S 2 followed by a brief discussion of the numerical scheme \S 2.1. Various aspects of individual test cases and the general evolution of the shocked cloud are discussed respectively in \S 3, and \S 3.1. The results are presented in \S 4 with conclusions in \S 5.

\begin{figure}
 \vspace*{4pt}
  \includegraphics[angle=0,width=8.cm]{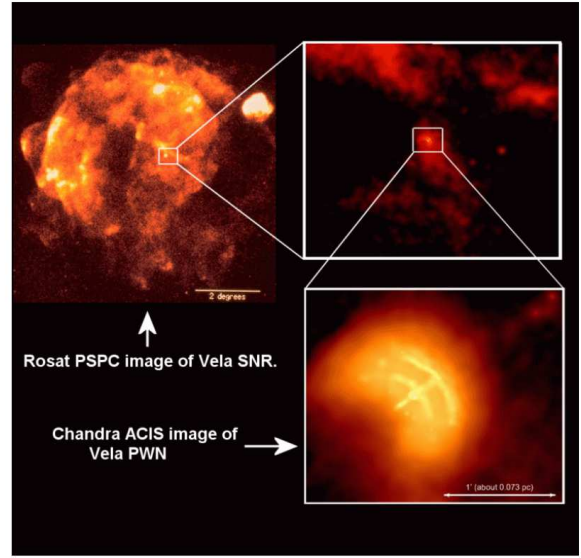}
 \caption{The Vela SNR photographed by the \emph{ROSAT} shows several bright regions, and an outflow has been detected in one of them. The adjoining boxes show respectively, the region where the outflow has been detected, and the Chandra image of the outflow. Image adopted from the electronic Chandra gallery (\emph{http://chandra.harvard.edu/}).}
\end{figure}

\begin{table*}
 \centering
 \begin{minipage}{100mm}
  \caption{Summarising physical details of the simulations discussed in the \newline paper. The artificial viscosity parameters $(\alpha,\beta)$ have also been specified.}
  \begin{tabular}{@{}lll@{}}
  \hline 
 Serial & Mass of \footnote{\tiny{$m_{i},m_{j},m_{k}$}-belonging, respectively to the cloud, the ICM, and the slab} & Physical parameters\footnote{\tiny{$M_{cld}, R_{cld}, T_{cld}, v_{cld}, \bar{\rho}_{cld}$-respectively the mass of the cloud, its radius, temperature, precollision velocity, \\ and the average density}}  \\
number & individual particle (M$_{\odot}$)& of the cloud \\
\hline
& {\small ISOTHERMAL SHOCK} \\
& $(\alpha, \beta)\equiv$(0.1,0.2)\\
\hline
1 & $m_{i}$=1.74$\times10^{-2}$ & $M_{cld}$=500 M$_{\odot}$, $R_{cld}$=0.5 pc \\
  & $m_{j}$=1.74$\times10^{-5}$ & $\bar{\rho}_{cld}\sim 6\times10^{-20}$ g cm$^{-3}$, \\
&$m_{k}$=3.67$\times10^{-5}$ & $T_{cld}$=15K\\
\hline
2 & $m_{i}$=1.74$\times10^{-4}$ & $M_{cld}$=5 M$_{\odot}$, $R_{cld}$=0.5 pc \\
  & $m_{j}$=1.74$\times10^{-6}$ & $\bar{\rho}_{cld}\sim 6\times10^{-22}$ g cm$^{-3}$, \\
 & $m_{k}$=2.06$\times10^{-6}$ & $T_{cld}$=80K\\
\hline
3 & same as in case 1 & same as in case 1 \\
  & & $\sigma_{cld}$=0.91 km s$^{-1}$, $\alpha_{vir}\sim$ 1\footnote{\tiny{$\alpha_{vir}$-Virial coefficient of the precollision cloud}}\\
\hline
4 & same as in case 2 & same as in case 2 \\
 & (ADIABATIC SHOCK)\\
\hline
&  $(\alpha, \beta)\equiv$(1,2)\\
5 & same as in case 1 & same as in case 1 \\
& (ISOTHERMAL SHOCK) \\
\hline
& HIGHER RESOLUTION \\
6 & $m_{i}$=2.20$\times10^{-3}$ & same as in case 1 \\
  & $m_{j}$=2.53$\times10^{-6}$ & \\
  & $m_{k}$=1.26$\times10^{-5}$ & \\
\hline
 & WITHOUT ARTIFICIAL CONDUCTIVITY \\
7 & same as in case 1 & same as in case 1 \\
\hline
& LOW RESOLUTION \\
8 & $m_{i}$=2.93$\times10^{-2}$ & same as in case 1 \\
  & $m_{j}$=2.93$\times10^{-5}$ & \\
  & $m_{k}$=1.09$\times10^{-4}$ & \\
\hline
\end{tabular}
\end{minipage}
\end{table*}

\section[]{Initial conditions}
Figure 2 shows a schematic representation of the cloud-shell system. To compensate for our limited computational resources, we consider only a section of the shell, modelled as a thin, dense slab having temperature, $T_{shell}$, and moving at velocity, $V_{s}$. We admit that such a choice for the initial conditions will reduce the post-collision momentum delivered to the stationary cloud.
The molecular cloud (MC) is modelled as a uniform density sphere, confined by a diffuse, warm intercloud medium (ICM) [$\lesssim$1 cm$^{-3}$; and a few times 10$^{4}$ K]; pressure equilibrium is maintained at the cloud-ICM interface. In {\small SPH}, the ICM is represented by special type of particles that exert only hydrodynamic force on ordinary gas particles. The entire assembly is enclosed in a periodic-box that is used to ghost particles, i.e. particles leaving from one face of the box arrive from the opposite face. The simulations were terminated once the slab traversed the box-width. Six test cases had $\sim$400,000 particles with $\sim$30,000 particles in the cloud, and $\sim$80,000 particles in the slab; the rest were ICM particles. The average {\small SPH} smoothing length, $h_{avg}$, within the cloud is
\begin{displaymath}
h_{avg}\sim\frac{1}{2}\Big(\frac{N_{neibs}}{N_{cld}}\Big)^{1/3}R_{cld},
\end{displaymath}
$N_{cld}$=30,000; for number of neighbours of an {\small SPH} particle, $N_{neibs}$=50, $h_{avg}\sim$ 0.03 pc $< \lambda_{J}=(\pi a_{0}^{2}/G\bar{\rho})^{1/2}\sim$ 0.2 pc, the Jeans length for gas at 20 K and average density in the shocked cloud, $\bar{\rho}\sim$ 10$^{-19}$ g cm$^{-3}$. We repeated the calculations for one of the test cases, case 1 listed in Table 1, with an eightfold increase in the number of particles within the cloud alone, with $\sim$240,000 particles in the slab, and 730,000 particles in the ICM so that the computational domain collectively had 1.2$\times 10^{6}$ particles. Then for the same of the number of neighbours, $N_{neibs}$, $h_{avg}\sim 1.5\times 10^{-2}$ pc, and $h_{avg}/\lambda_{J} \sim 13$, sufficient to resolve the Jeans instability according to the Truelove criterion (Truelove \emph{et al.} 1998). For an arbitrarily chosen mass of the cloud, $M_{cld}$, its radius, $R_{cld}$, is deliberately chosen much smaller than that prescribed by the Larson's relation
\begin{equation}
R_{cld}(\textrm{pc})=0.1(\textrm{pc})\Big(\frac{M_{cld}}{\textrm{M}_{\odot}}\Big)^{0.5},
\end{equation}
\citep{b3}. The sound-speed, $a_{0}$, inside a self-gravitating cloud must be much smaller than its escape velocity, $v_{g}$, defined as, $v_{g}^{2}=\frac{2GM_{cld}}{R_{cld}}$. The temperature within the cloud, $T_{cld}$, is chosen such that $a_{0}^{2}\ll v_{g}^{2}$. The turbulent velocity field in the MC in case 4 is set up in a way similar to that described by \citet{b29}. The velocity perturbations are so set that the initial power spectrum is $P(k)\propto k^{\alpha}$, $\alpha=0$; the initial amplitude of the velocity perturbations is set using the velocity scaling relation due to \citet{b3}. The physical parameters for each test case have been listed in Table 1 above.

\subsection{Smoothed particle hydrodynamics {\small(SPH)}}

The simulations discussed here have been performed using {\small SPH}, a Lagrangian, particle based scheme. An {\small SPH} particle, contrary to intuition, has a finite size characterised by its smoothing length, $h$. Thus, an {\small SPH} particle carries physical properties such as mass, density, and velocity; {\small SPH} particles interact with each other via numerical viscosity, called artificial viscosity (AV). In all test cases, but case 5, we have used the attenuated form of viscosity prescribed by \citet{b4}, that has been demonstrated to better represent shocks. Under the scheme {\small SPH} viscosity parameters $(\alpha,\beta)\equiv$ (0.1,0.2). In addition, we also use the Artificial thermal conductivity scheme devised to smooth out discontinuities across regions with steep density gradients \citep{b39}. In the fifth case we have employed the conventional {\small SPH} viscosity, with $(\alpha,\beta)\equiv$ (1,2), combined with Artificial conductivity. This is intended to demonstrate the detrimental effect of {\small SPH} viscosity on shocks, and the formation of post-shock structure through the interplay of various dynamical instabilities. We used the well tested {\small SPH} code, Seren, for our purpose here. The robustness of the code has been demonstrated by the authors through a suite of tests, that includes modelling hydrodynamic instabilities such as the Kelvin-Helmholtz instability (KHI) and the non-linear thin-shell instability (NTSI) (Hubber \emph{et al.} 2010 a,b).
The code uses a Barnes-Hut tree \citep{b2}, to find the nearest neighbours of an {\small SPH} particle and to calculate the net force on it. Seren also includes the quadrupole moments of distant cells in its calculation of self-gravity. Seren employs the modified M4 kernel suggested by Thomas \& Couchman (1992).

As in our previous work \citep{b1}, thermodynamic details in the first four cases here have been tackled using a simple barotropic equation of state (EOS), defined by Eqn. (2) below, and plotted in Fig. 3, while the post-collision shock in case 5 is purely adiabatic, defined by the third part of Eqn. (2). (The adiabatic gas index, $\gamma= 5/3$).

\begin{equation}
\frac{P}{\rho} = (k_B/\bar{m})\times
\left\{ \begin{array}{ll}
\Big(\frac{T_{shell}}{\textrm{K}}\Big)\ ; \rho\le10^{-23} \textrm{g cm}^{-3}\\
\Big(\frac{T_{cld}}{\textrm{K}}\Big)\ ; 10^{-23}\textrm{g cm}^{-3} <\rho
\le 10^{-22}\textrm{g cm}^{-3}\\
\Big(\frac{\gamma T_{cld}}{\textrm{K}}\Big)\Big(\frac{\rho}{10^{-22}\textrm{g cm}^{-3}}\Big)^{\gamma-1}; 10^{-22}\textrm{g cm}^{-3}\\
< \rho\le 5\times 10^{-22}\textrm{g cm}^{-3}\\
\Big(\frac{T_{cld}}{\textrm{K}}\Big)\ ; 5\times 10^{-22}\textrm{g cm}^{-3}<\rho \\
\le 10^{-18}\textrm{g cm}^{-3}\\
\Big(\frac{T_{iso}}{\textrm{K}}\Big)\Big[1 + \gamma\Big(\frac{\rho}{10^{-15}\textrm{g cm}^{-3}}\Big)^{\gamma-1}\Big]\ ; \\
\rho > 10^{-18}\textrm{g cm}^{-3}.
\end{array} \right.
\end{equation}
The first two parts of the EOS are designed to maintain the suitable temperatures in the shell, and the cloud, respectively. The immediate post-shock temperature within the MC is allowed to rise adiabatically using the third part of the EOS after which the gas is cooled down to its precollision temperature, $T_{cld}$, as can be seen from the fourth part of the EOS. The gas downstream of the shock, after becoming sufficiently dense, is cooled further and maintained at temperature, $T_{iso}$ = 10K. The last part of the EOS, that allows the temperature to rise adiabatically when condensation occurs, is never broached in these simulations.

\begin{figure}
 \vspace*{5pt}
  \includegraphics[angle=0,width=8.cm]{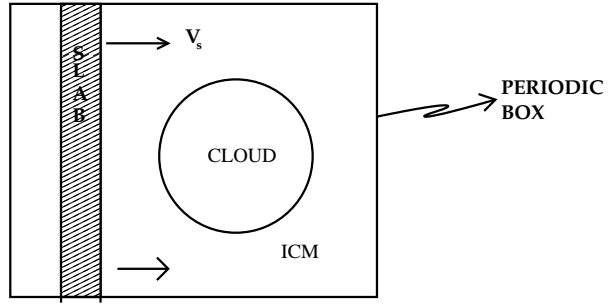}
 \caption{Schematic representation of the cloud-shell system in the plane of collision. A section of the expanding shell, shown as a plane slab, moves towards a stationary cloud with a velocity $V_{s}$. Note that the cloud is embedded in a warm intercloud medium.}
\end{figure} 

\begin{figure}
 \vspace*{5pt}
  \includegraphics[angle=270,width=7.5cm]{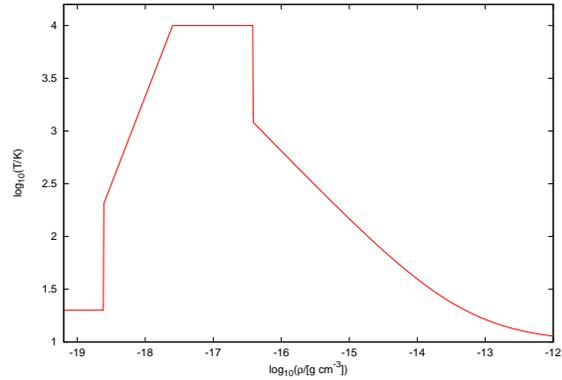}
 \caption{The EOS defined by Eqn. (2) above in the logarithmic $T-\rho$ space.}
\end{figure} 

\begin{figure}
 \vspace*{5pt}
  \includegraphics[angle=270,width=7.5cm]{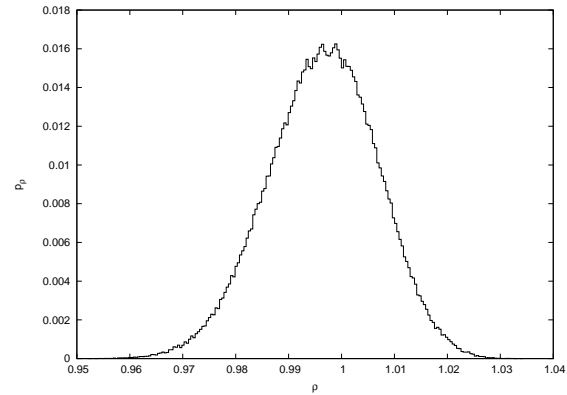}
 \caption{A plot demonstrating the stability of the preshock cloud in isolation. It shows the probability distribution, $p_{\rho}$, of the density of {\small SPH} particles in a relaxed cube of unit dimensions. The distribution peaks at unity, the true density of the box.}
\end{figure} 

\section{Evolution of the shocked cloud } 

\subsection{Stability of the precollision cloud }
Before proceeding to discuss the evolution of the shocked cloud, we wish to demonstrate that the observed structure in the shocked cloud is indeed physical, produced due to propagation of shock in the cloud. The initial conditions for the numerical experiment were set up using a relaxed configuration of particles distributed uniformly in a cube of unit length. Particles in the cube were relaxed by evolving the box for about 5 sound-crossing times so that its average density is within a few percent of its true value, $\rho_{true}$. Figure 4 shows the probability distribution for the density of {\small SPH} particles in the relaxed box, which as expected, peaks near $\rho_{true}$. The required uniform density cloud was scooped out of this relaxed box that does not show evidence of clumping i.e. spurious density enhancements, over the relevant dynamical timescale, the crossing time, $t_{c}$, for the cloud, defined by Eqn. (3) above.
 
\subsection{Summary }
 Although the set of test cases discussed in this paper could possibly be divided into 3 subgroups as indicated in Table 1 above, the general nature of evolution of the shocked cloud in all cases is rather similar. The precollision cloud in each case is colder and denser than the incident slab, and so, as a direct consequence of the equation of continuity, it follows that the post-collision shock would propagate within the cloud at a much lower velocity. This, indeed is observed in the simulations and the shock-wave propagating within the cloud forms a wake, a bow like structure. The impact of the incident slab on the cloud surface reflects a shock wave in the ICM, as is evident from the rendered density plots of Fig. 5, and the supplementary animation. These plots show the cross section of the cloud taken through its mid-plane and the density is measured in M$_{\odot}$pc$^{-3}$; time on all rendered plots is marked in Million yrs. Gas shearing the cloud surface renders it unstable to the Kelvin-Helmholtz (KH) instability, while the cloud accelerated by the shock-wave within its interiors renders it Rayleigh-Taylor (RT) unstable. The incident slab, due to its finite thickness losses momentum after shocking the ICM and the cloud, and consequently the pressure at the rear end of the cloud is insufficient to trigger its implosion. So, despite our observation in the present work being apparently contradictory to that reported previously by, for instance \citet{b36}, and \citet{b37}, in fact, there is none; the present work effectively tackles the case of a thin shell impacting a cloud which is clear from the argument below in \S 3.3. We argue that shock-induced turbulence generates dense structure on a rather short timescale, some of which could become self-gravitating and go on to produce prestellar cores. This result holds irrespective of the nature of shock, as can be seen from the ,comparative plots in Fig. 6. Convergence of results is tested by repeating case 1 (\S 4.2) with eight times the number of particles in the preshock cloud and then at the other extreme, with approximately half the number of particles; see Fig. 12 below. Next, we discuss our results in some more detail.

\begin{figure*}
\vbox to 65mm{\vfil 
  \includegraphics[angle=270,width=18.cm]{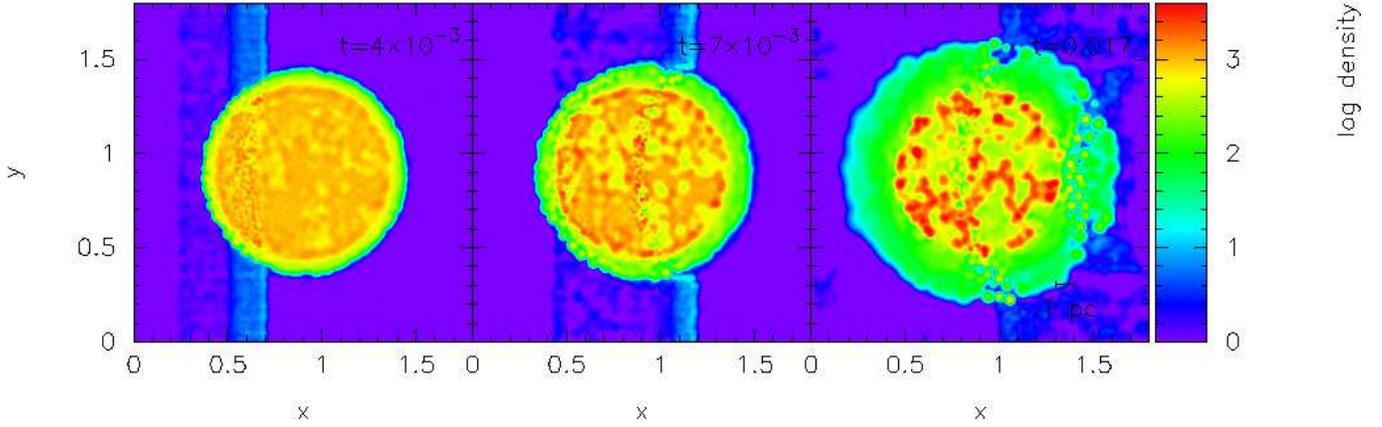}
\caption{A scroll showing the mid-plane, cross-sectional rendered density plots (measured in M$_{\odot}$pc$^{-3}$) of the shocked cloud in case 1. The high velocity slab impinging on the cloud from the left drives a shock-wave into the latter and injects turbulence within the cloud; shortly thereafter, ($t=0.017$ Myrs), turbulence produces a network of filaments, much denser than the precollision cloud. \emph{Animation available for online version only.}} \vfil} 
\end{figure*}

\begin{figure*}
\vbox to 92mm{\vfil 
  \includegraphics[angle=270,width=18.cm]{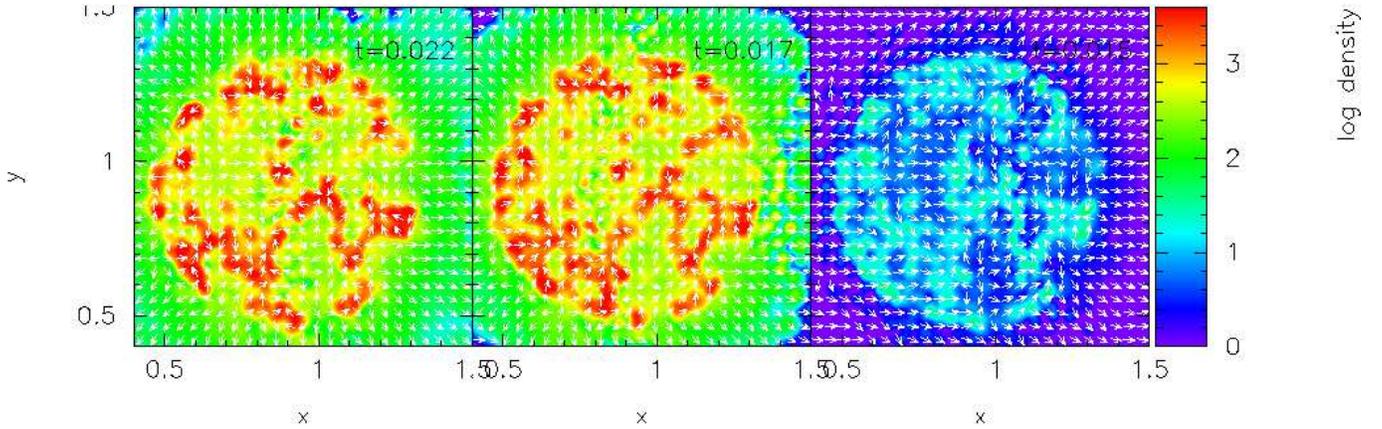}
\caption{From left to right, the panels show a rendered cross-sectional density plot of the shocked cloud on termination of the simulations in cases 1, 3, and 4, listed in Table 1 above. It appears that an additional turbulent velocity field assists structure formation, as is evident from the comparatively shorter 
timescale on the plot in the central panel, relative to that on the left-hand panel. The plot in the right hand panel, that of an adiabatic shock, shows less fragmentation compared to the earlier two cases. \emph{See text for description}.} \vfil} 
\end{figure*}

\subsection{Dynamical evolution of the shocked cloud}
Before embarking upon a discussion of various test cases, it will be useful to define some timescales which will be summoned in further analysis. A slab moving with a velocity, $V_{s}$, would traverse across a cloud of radius, $R_{cld}$, in a crossing time, $t_{c}$, defined as 
\begin{equation}
t_{c}\sim\frac{2R_{cld}}{V_{s}}.
\end{equation}
The cloud crushing time, $t_{cc}$, the timescale over which a shocked cloud might flatten is,
\begin{equation}
t_{cc}\sim\frac{\chi^{1/2}2R_{cld}}{V_{s}},
\end{equation}
where $\chi=\frac{\rho_{cld}}{\rho_{ICM}}\sim$1000, in all the simulations discussed. 

Of greater importance is the dynamical timescale, $t_{d}$, of the incident shock, the timescale over which pressure behind the incident shock may vary. When $t_{cc}\ll t_{d}$, the shock is practically unaffected even after impacting the cloud. On the contrary, when $t_{d}\ll t_{cc}$, the dynamical properties of the shock change substantially after impacting the cloud. Thus, a qualitative definition of a small and a large cloud may be adduced to these preconditions: a cloud in which $t_{cc}\ll t_{d}$, may be classified as small, while a large cloud is one with $t_{d}\ll t_{cc}$. For illustrative purposes we shall adopt the connotation of denoting the preshock state variables with the subscript '1', and corresponding post-shock variables with '2'. The post-shock velocity of the slab can be obtained simply from the equation of continuity as,
\begin{equation}
V_{s2} = \Big(\frac{\rho_{cld1}}{\rho_{cld2}}\Big)V_{s1}.
\end{equation}

The pressure, $p_{cld2}$, in the shocked cloud is given by the pressure-jump condition as,
\begin{equation}
\frac{p_{cld2}}{p_{cld1}}=\frac{2\gamma\mathcal{M}^{2}-(\gamma-1)}{(\gamma+1)},
\end{equation}
$\mathcal{M}$ is the precollision Mach number (Shore 2007). For a post-shock temperature, $T_{2}$, the corresponding sound speed, $a_{2}$, is related to the pressure simply by,
\begin{equation}
\frac{p_{cld2}}{p_{cld1}}=\frac{a_{2}^{2}\rho_{cld2}}{a_{0}^{2}\rho_{cld1}}.
\end{equation}

The maximum post-shock density for a radiative shock is $\rho_{cld2}\sim\mathcal{M}^{2}\rho_{cld1}$; so that combining this condition with Eqns.(5), (6), and (7), we get,
\begin{equation}
\mathcal{M}^{2}=\Big(\frac{p_{cld2}}{p_{cld1}}\Big)\Big(\frac{a_{0}}{a_{2}}\Big)^{2}=\frac{V_{s2}}{V_{cld2}},
\end{equation}
also for a strong shock, it follows from Eqn. (6),
\begin{equation}
\frac{p_{cld2}}{p_{cld1}}\sim 2\gamma\mathcal{M}^{2}.
\end{equation}
Equations (8) and (9) led us to-
\begin{displaymath}
 2\gamma\Big(\frac{a_{0}}{a_{2}}\Big)^{2} =  \frac{V_{s2}}{V_{cld2}}.
\end{displaymath}

Now, for a shock that has traversed the cloud radius, and remained fairly stable, $R_{cld}\sim t_{d}v_{cld2}$, so that from the equation above we have:
\begin{equation}
t_{d}\sim\frac{R_{cld}}{V_{s2}}2\gamma\Big(\frac{a_{0}}{a_{2}}\Big)^{2}.
\end{equation}
As a direct consequence of the temperature-jump condition, we have
\begin{displaymath}
\Big(\frac{a_{0}}{a_{2}}\Big)^{2} = \frac{(\gamma+1)^{2}}{2\gamma\mathcal{M}^{2}(\gamma-1)},
\end{displaymath}
so that Eqn. (10) finally becomes, 
\begin{equation}
t_{d}\sim\frac{R_{cld}}{V_{s2}}\frac{(\gamma+1)^{2}}{\mathcal{M}^{2}(\gamma-1)}.
\end{equation}
In the present work, with $\gamma=5/3$, and $\mathcal{M}\sim$ 25, so
\begin{equation}
t_{d}\sim\frac{\eta R_{cld}}{V_{s2}},
\end{equation}
where $\eta\sim$0.1.

\textbf{Cases 1 and 3 (Velocity of the incident shock, $V_{s}$ = 200 km s$^{-1}$, $\frac{\rho_{cld}}{\rho_{s1}}\sim$2) :} \\
The front surface of the slab, on colliding with the stationary cloud generates a shock wave in the latter, and after a brief interval of time there is a second, weaker shock when the trailing end of the incident slab shocks the cloud. This double shock is evident from the step in the red curve of the top left-hand panel of Fig. 8 below. The initial impact of the slab reflects a rarefaction wave in the ICM at the periphery of the cloud, characterised by the leftward shift of the green curve relative to the red curve, and a dip in the former near the front-edge of the cloud, which after being shocked, has now become denser. This is also evident from the plot in the central panel of Fig. 5, in which the denser edge of the shocked cloud is visible. The shock propagating within the cloud however, is progressively weakened as evidenced by the progressive drop in pressure within the cloud at latter epochs, shown in the top left-hand panel of Fig. 8.

The propagating shock generates turbulence with the cloud which itself is dissipative, and produces fractal structure which at latter epochs, appears more filamentary with a few clumps. The left-hand panel of Fig. 6 shows the fragmented interiors of the shocked cloud, the velocity vectors overlaid on the plot show the underlying chaotic field. Apparently, this structure develops on a timescale comparable to the crossing time, $t_{c}$, defined by Eqn. (3) above. The length of the fastest growing mode, $\lambda_{turb}\sim\Big(\frac{\pi v_{eff}^{2}}{G\bar{\rho}}\Big)^{1/2}$, and $v_{eff}\sim (a_{0}+v_{cld2})$; $v_{cld2}\sim V_{s1}/\mathcal{M}^{2}\sim$ 0.4 km s$^{-1}$, $a_{0}\sim$ 0.27 km s$^{-1}$ at 20 K.
 Then for a typical post-shock density, $\bar{\rho}\sim 10^{-18}$ g cm$^{-3}$, $\lambda_{turb}\sim$ 0.1 pc, an order of magnitude smaller than $h_{avg}$ for this case. Thus the fragmentation observed in the shocked cloud here is likely to be physical. 

Following a shearing interaction between the slab and the outer layers of the cloud, its surface is rendered KH unstable. Although, here we are able to see only a few blobs, broken away from the cloud surface; see for instance the last panel in Fig. 6. This could perhaps be the case since the shearing surface in these test cases is not spatially extensive, following which there is not sufficient shearing contact. The thin slab surface in contact with the main body of the cloud, however, shows a few rolls, but these are not particularly well developed as can be seen from the density plot, with overlaid contours shown in Fig. 7. Another possible reason could be the inability of {\small SPH} to handle density contrasts within fluid layers, which is well documented in contemporary literature. While remedies have been suggested to alleviate some of the problems related to fluid mixing, their robustness in presence of self-gravity is untested. We shall discuss one such exploration in case 7 below, where the issue of resolving the KH instability in this work will be briefly revisited.

In case 3, we use the same initial conditions as in case 1, but the precollision cloud in this case is additionally supported by a turbulent velocity field. The amplitude of the turbulent velocity is $\sim$0.9 km s$^{-1}$, so that the Virial parameter,$\alpha_{Vir}$, for the precollision cloud is approximately unity. This additional velocity field makes little difference to the evolution of the shocked cloud. The central panel of Fig. 6 shows the fragmented interiors of the cloud, from which it is evident that apart from the timescale of fragmentation, there is little change otherwise. The cloud in this case appears to fragment much faster and develops well-defined structure at an epoch somewhat earlier than in case 1. The two cases can be compared from the respective plots in the central, and the left-hand panel of Fig. 6. It therefore appears that turbulence aids formation of structure via fragmentation. \\ \\

\begin{figure*}
 \vspace*{5pt}
  \includegraphics[angle=270,width=6.5cm]{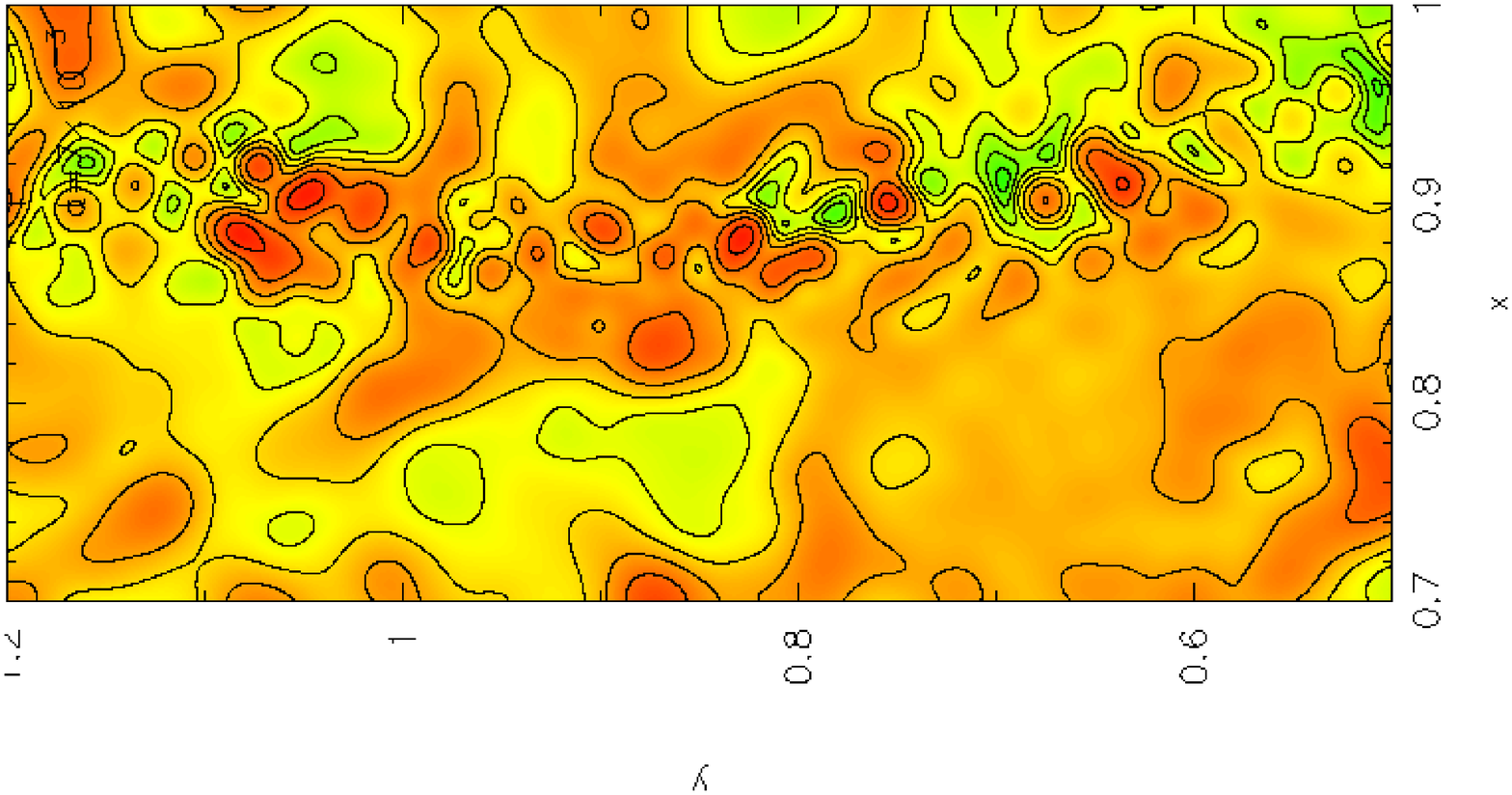}
  \includegraphics[angle=270,width=9.0cm]{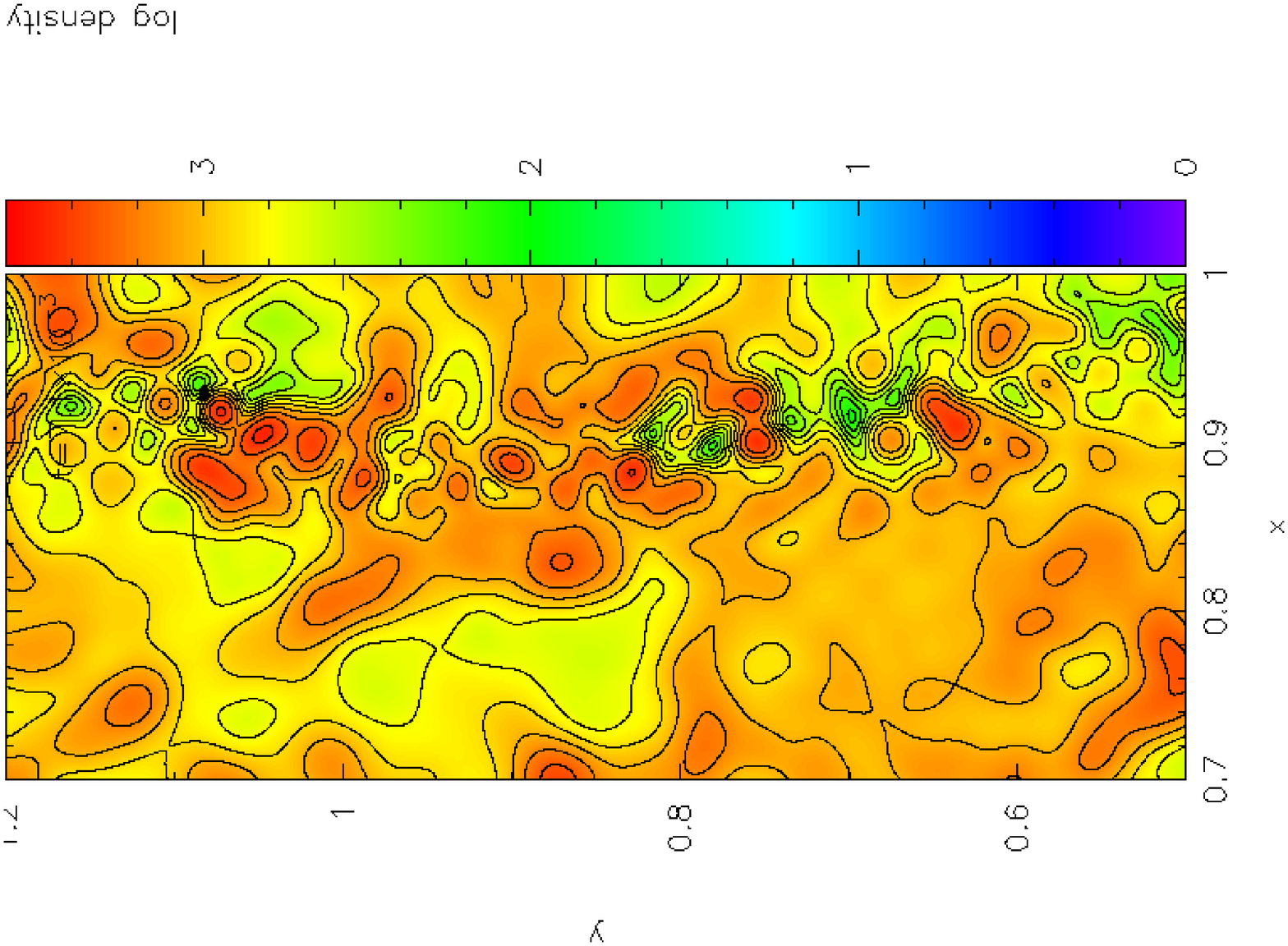}
 \caption{A coeval, cross-sectional plot comparing the growth of KH instability (KHI) due to shear between the incident slab and the cloud surface, with, and without the Artificial conductivity (AC). The plot in the right-hand panel is the one without AC, and apparently there is not much difference in the two cases. We do not observe the rolls associated with KHI; density contours have been overlaid on the rendered image to aid identification of underlying structure.}
\end{figure*}

\textbf{Cases 2 and 4 (Velocity of the incident shock, $V_{s}$ = 200 km s$^{-1}$, $\frac{\rho_{cld}}{\rho_{s1}}\sim$2) :}
The precollision cloud in case 2 is much more rarefied ($\sim10^{-23}$ g cm$^{-3}$) as compared to that in case 1, with the velocity of the incident slab remaining unchanged. The quantitative details of the post-collision cloud though, remain identical to those of case 1. The initial conditions of this case are also used for the fourth case. Having assumed a strongly radiative shock in the earlier cases, we adopted an adiabatic EOS in the fourth case, rendering the post-collision shock, non-radiative. Consequently the shocked gas ($\sim 10^{-21}$ g cm$^{-3}$) remains warm at a few hundred Kelvin. We notice that the warm gas, as expected, shows lesser propensity towards fragmentation and therefore greater contiguity in structure. The reader is referred to the right-hand panel of Fig. 6. It is unlikely for such fragments to be bounded and therefore, may simply be transitional features. 

A common feature of all these simulations is the relatively small value of the dynamical timescale $t_{d}$, defined by Eqn. (12) above. We observe that in all the test cases $t_{d}\ll t_{cc}$, so that pressure behind the slab changes even before it traverses across the cloud; consequently the incident slab is devoid of compressive strength by the time it reaches the rear end of the cloud, and so the cloud does not flatten. The timescale, $t_{d}$, calculated using Eqn. (12) above is $\sim10^{-4}$ Myr $\ll t_{cc}\sim$ 0.2 Myr. The problem, in the form tackled here, therefore invariably reduces down to a shock interacting with a large cloud or alternatively, a weak shock impacting a cloud. 

We are, however, unable to follow further evolution of the filaments and clumps since the periodic box enclosing the system has insufficient dimensions, and expansion of the reflected shock in the ICM, is impermissible; for mixing between opposite sides of the ICM shock must be avoided. The simulations were therefore terminated after the formation of dense structure. \\ \\
\textbf{Case 5 :}
While adopting the normal prescription of the {\small SPH} viscosity, the initial conditions were kept identical to those in the first. Not only does the higher viscosity strongly decelerate the incident slab, it also dissipates energy within the slab layers and in fact, renders it unstable via the growth of bending modes. Figure 9 shows a rendered, cross-sectional density plot of the slab-cloud system in this case. The wiggles on the slab surface appear similar to the features of the thin shell instability, but the initial conditions were set up such that the smallest unstable mode was much larger than the breadth of the computational box. This wavenumber can be calculated using Eqn. (14) deduced by \citet{b40}, for a thin shocked-shell. So these wiggles are possibly a manifestation of excessive dissipation due to stronger viscosity, and in fact, the viscous interaction renders the shock within the cloud, stationary, as is visible from the central and right-hand panels of Fig. 9.  

This is also the case in which the shocked cloud shows the least fragmentation, and therefore, the structure within the cloud is by and large contiguous. The fastest growing mode, $\lambda_{turb}$, in this case will be considerably shortened as the contiguous structure becomes denser which is likely to produce a large number of small fragments. This case demonstrates that a numerical study of hydrodynamical instabilities is likely to be corrupted by strong viscosity, and potentially misrepresent the situation. We defer the discussion of cases 6, 7, and 8 to \S 4.2, following that of the generation of vortices in the shocked cloud in \S 4.1.

\begin{figure*}
\vbox to 105mm{\vfil 
  \includegraphics[angle=270,width=7.cm]{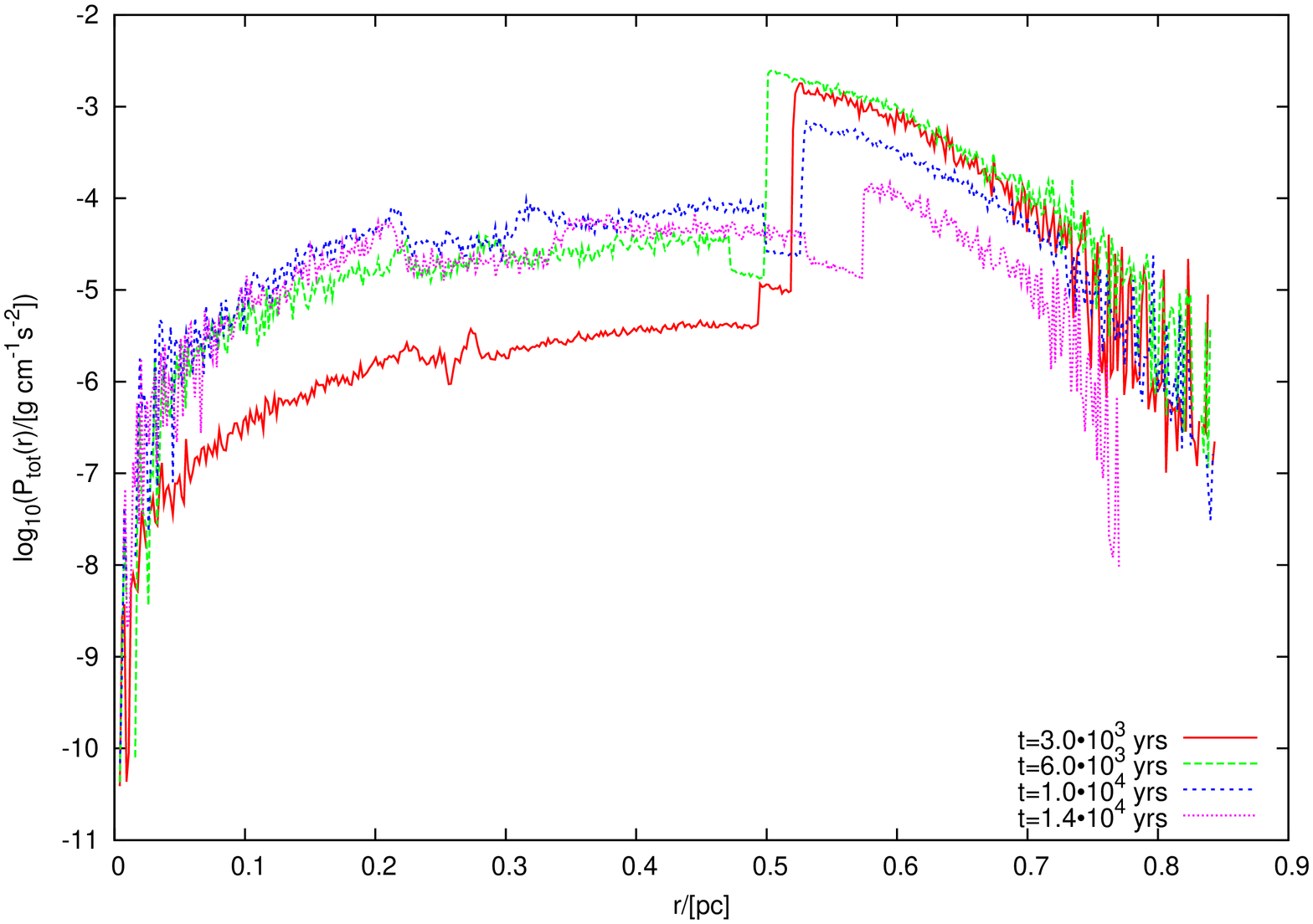}
  \includegraphics[angle=270,width=7.cm]{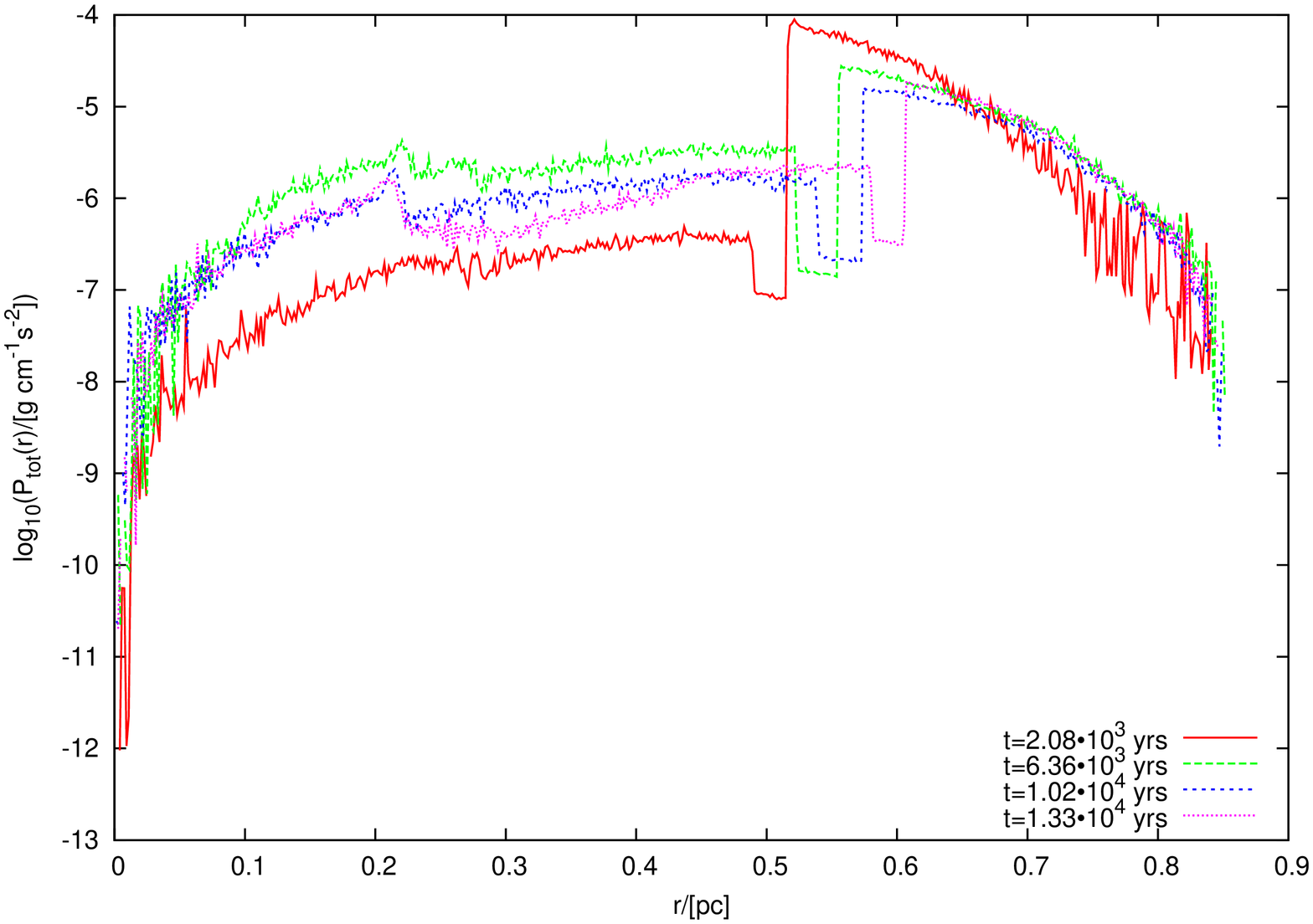} 
  \includegraphics[angle=270,width=7.cm]{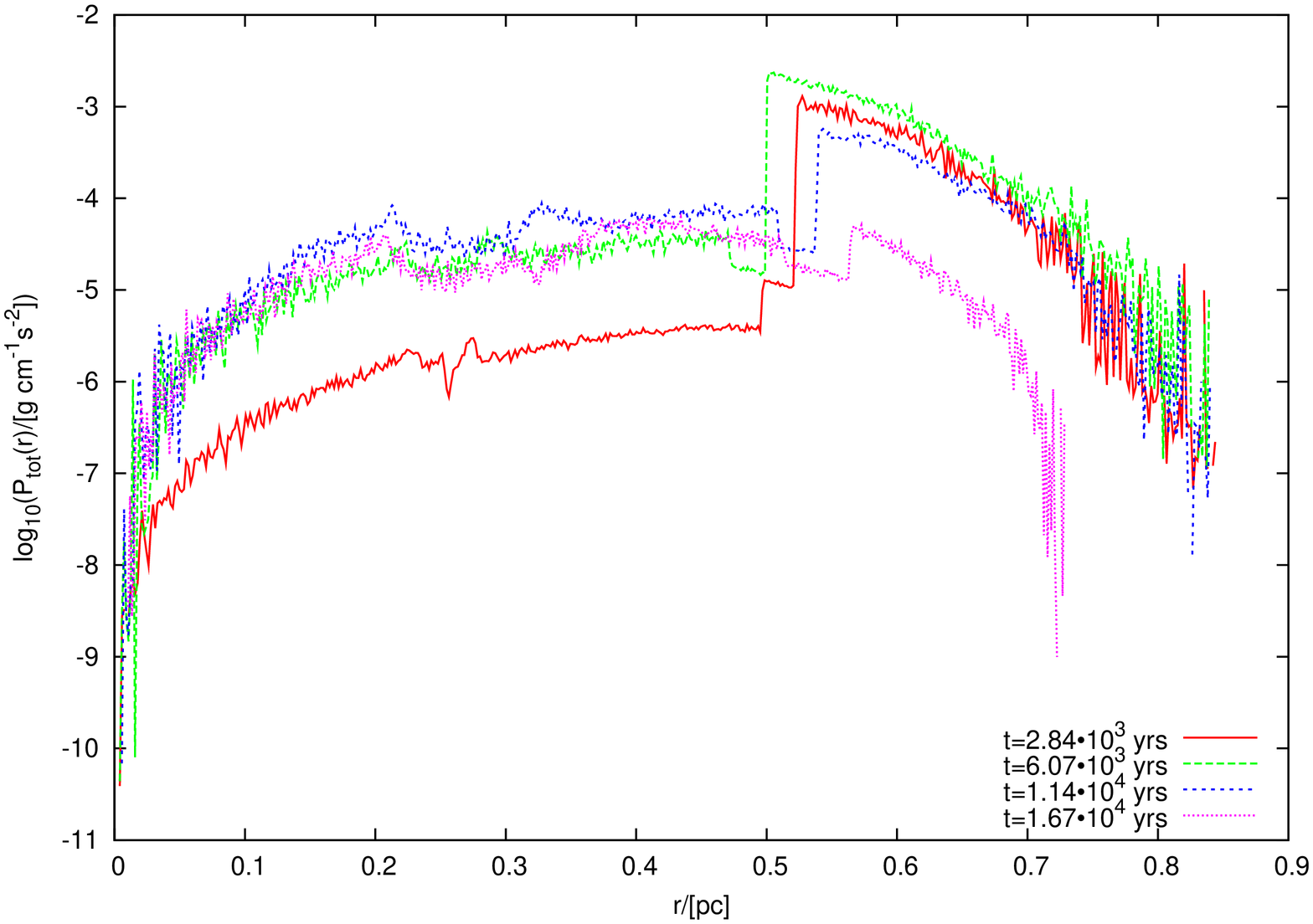}
  \includegraphics[angle=270,width=7.cm]{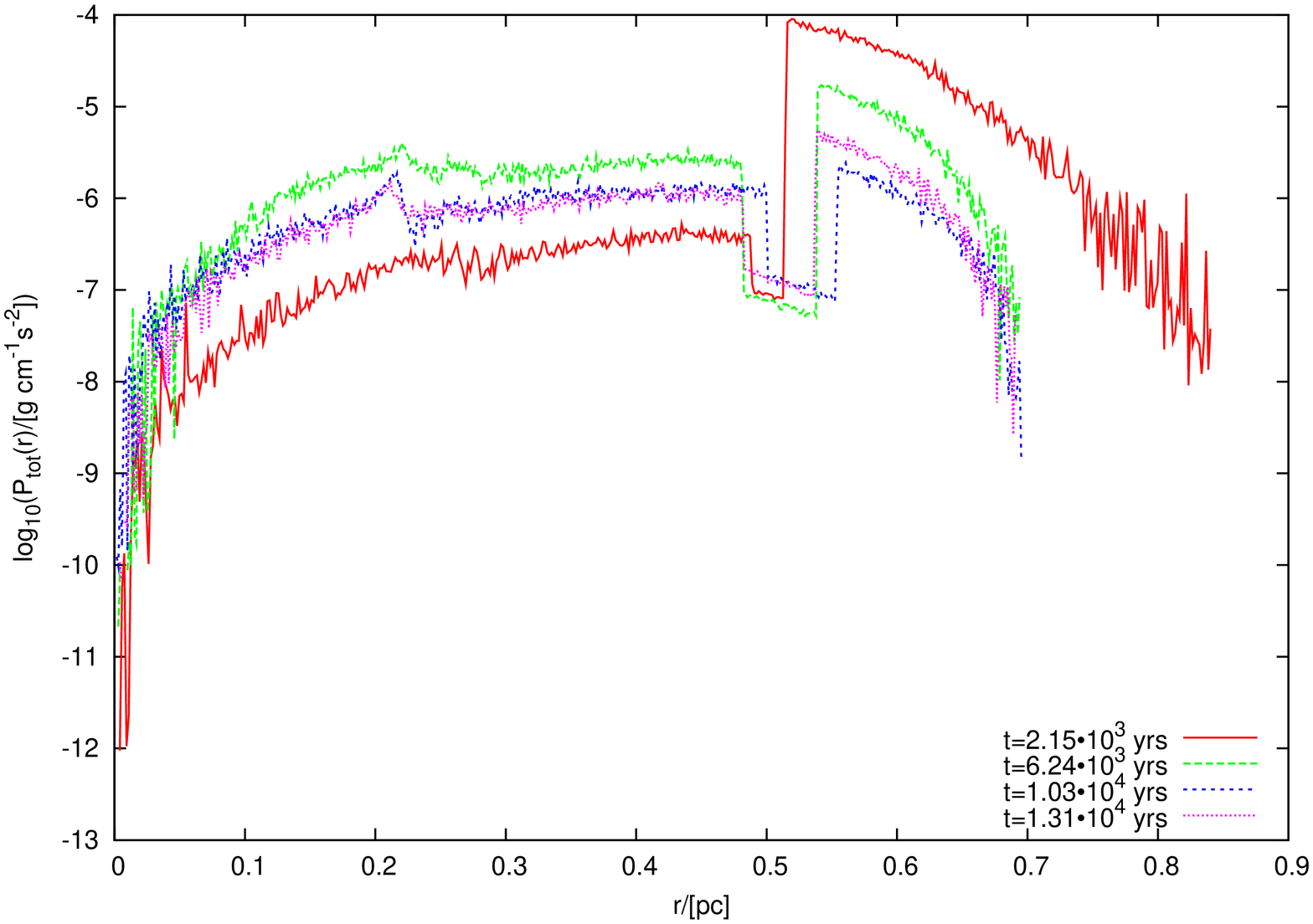} 
\caption{ Radial distribution of total pressure within the shocked cloud in the first 4 test cases; cases 1 \& 2 are plotted in the top left and right panels while 3 \& 4 are plotted in the bottom left and right hand panels. The plots have been made at different epochs of time as the post-collision shock propagates in the cloud. The shock following the impact of the incident slab causes steep discontinuity in the pressure distribution at the cloud edges, $r\sim$ 0.5 pc in the plots here. Observe that unlike in cases 1 \& 3, there is no reflected shock at the cloud surface in cases 2 \& 4, where the precollision cloud is much more rarefied. \emph{See text for description.}}\vfil} 
\end{figure*}

\begin{figure*}
\vbox to 81mm{\vfil 
  \includegraphics[angle=270,width=16.cm]{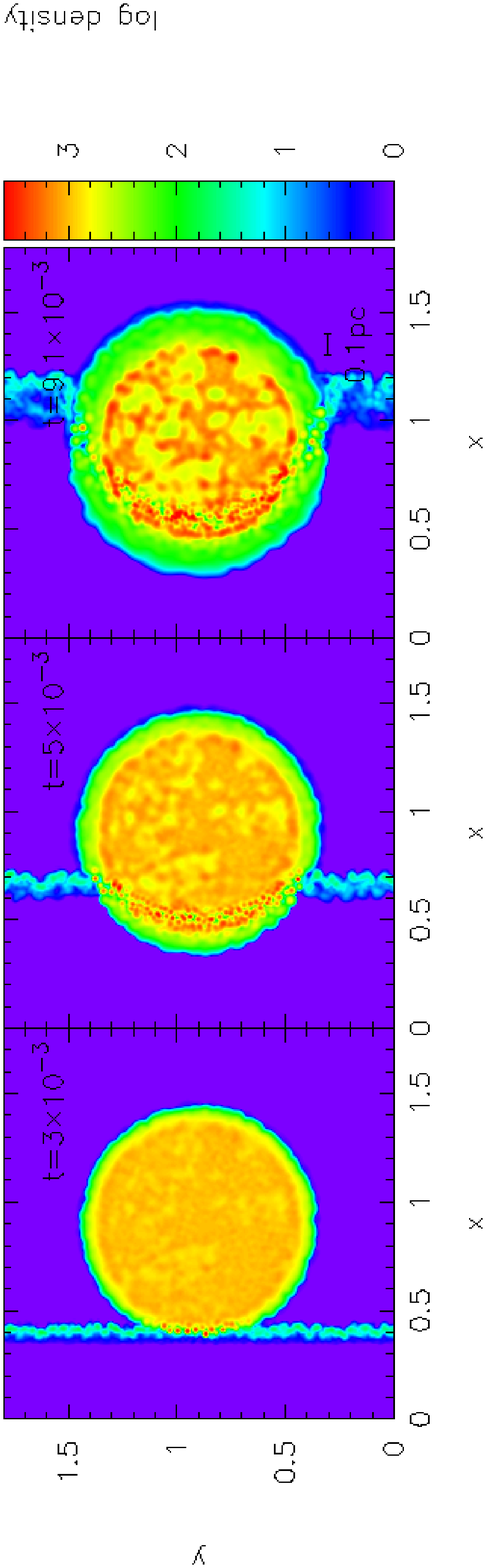}
\caption{A scroll showing the slab-cloud interaction in case 5. Notice that the incident slab itself, unlike in the first 4 cases, shows signs of instability and develops perturbations. Having impacted the cloud, the incident shock soon loses momentum and appears to have become stationary within the cloud; the wake in the shocked cloud shown in the right-hand panel is still at its front end. \emph{See text above.}} \vfil} 
\end{figure*}

\section{Discussion}

We have, in the preceding section demonstrated that the general evolution of a shocked cloud is rather insensitive to the nature of the post-collision shock. Although the extent of fragmentation in the shocked cloud appears to depend on the initial conditions, and the post-shock cooling. These results are radically different from those reported by \citet{b36}, it must, however, be noted that his work and indeed, a latter work by \citet{b38} with magnetised precollsion clouds neglected self-gravity. Self-gravity has been included in the present work, and the evolution of the shocked cloud apparently depends on the balance between the gravitational force per unit area and the ram-pressure due to turbulent motion.

\subsection{Vorticity in the post-shock medium}

An important point demanding consideration is that of vorticity generated in the post-shock cloud, and the surrounding ICM. It is well known that an ideal fluid is free of vorticity, the situation with the fluid in these simulations or any other numerical treatment differs in two ways : (i) the inevitable presence of numerical viscosity in the numerical interpolation renders the fluid non-ideal, and (ii) shocks in general are not isentropic. The importance of turbulence in structure formation and in general, on the process of star-formation via the process of turbulent fragmentation, is well appreciated. It is therefore quintessential to look in to the possible sources of turbulence. 

While an ideal fluid being curl-free, always remains so as a consequence of the Kelvin's vorticity theorem, this is hardly true in case of real fluids or those of the type considered here. In these latter fluids vorticity could be created, destroyed, or may simply diffuse from one part to another part of the fluid. To see this, let us go through the following set of simple equations: The vorticity, \boldmath{$\omega$}=$\nabla\times\textbf{v}$, for a fluid having velocity, $\textbf{v}$, the time derivative of which is,
\begin{displaymath}
\frac{\partial {\boldmath{\omega}}}{\partial t} = \nabla\times\dot{\textbf{v}}.
\end{displaymath}
The quantity \boldmath{$\omega\times \textrm{\textbf{v}} = \textrm{\textbf{v}}\cdot(\nabla \textrm{\textbf{v}})-\frac{1}{2}\nabla \textrm{\textbf{v}}^{2}$}, has a vanishing curl, so that the above expression may be re-written as,
\begin{equation}
\frac{\partial \boldmath{\omega}}{\partial t} + \boldmath{\nabla\times (\omega\times\textrm{\textbf{v}})} = \nabla\times\dot{\textbf{v}}.
\end{equation}
The equation of fluid motion including the viscosity is,
\begin{displaymath}
\frac{\partial \textrm{\textbf{v}}}{\partial t} = -\Big(\frac{1}{\rho}\boldmath{\nabla} p + \nu\nabla^{2}\mathrm{\textbf{v}}\Big) - \boldmath{\nabla}\phi_{g};
\end{displaymath}
where $\nu$ is the kinematic viscosity of the fluid, and $\phi_{g}$ is the gravitational potential. The remaining symbols have their usual meaning. Using this expression in the right-hand side of Eqn. (13), and following a little manipulation we wind up with,
\begin{equation}
\frac{\partial \boldmath{\omega}}{\partial t} + \boldmath{\nabla\times (\omega\times\textrm{\textbf{v}})} = -\Big[\boldmath{\nabla}p\times\boldmath{\nabla}\Big(\frac{1}{\rho}\Big) + \nu\nabla^{2}\boldmath{\omega}\Big],
\end{equation}
which is the equation describing the time evolution of the vorticity of the fluid. The second term on the right-hand side of Eqn. (14) above defines the spatial distribution of vorticity in the shocked gas. Now using the Laplacian identity, this term can be re-written as 
\begin{displaymath}
[\boldmath{\nabla}\cdot(\boldmath{\nabla}\cdot\boldmath{\omega}) - \boldmath{\nabla}\times\boldmath{\nabla}\times\boldmath{\omega}] = \boldmath{\nabla}^{2}\boldmath{\omega}
\end{displaymath}
The gradient of a curl is zero so that the first term on the left-hand side vanishes, and we are left with
\begin{displaymath}
-\boldmath{\nabla}\times\boldmath{\nabla}\times\boldmath{\omega} = -\boldmath{\nabla}\times\boldmath{\nabla}^{2}\textbf{v} = \boldmath{\nabla}^{2}\boldmath{\omega}
\end{displaymath}
The quantity $(\boldmath{\nabla}\times\boldmath{\nabla}^{2}\textbf{v})$ has been plotted for the shocked gas in the cloud, shown in Fig. 10. The shocked gas, in general, shows smaller vorticity which suggests, post-shock, vorticity is dissipated.

An useful quantity to study the temporal evolution of vorticity is its circulation, W, defined as 
\begin{equation}
\textrm{W} = \int_{S} \boldmath{\omega}\cdot d\textbf{S},
\end{equation}
$d\textbf{S}$ being an infinitesimally small area element. This quantity gives the flux of vorticity through an area element $d\textbf{S}$. We have plotted the time evolution of the net vorticity in cases 1 and 5. To remind our readers, we may recount that starting with identical initial conditions and the EOS, case 5 employs the conventional {\small SPH} viscosity, while the attenuated viscosity, i.e. the Morris-Monaghan prescription is employed in case 1. Figure 11 shows a comparative plot of the net vorticity for these two cases. It is clear from these plots that a larger {\small SPH} viscosity generates greater vorticity in the post-shock cloud of case 5, although the vorticity in either cases decays with time. However, a larger vorticity will lead to enhanced dissipation of kinetic energy within the system which in turn will have grave implications on its dynamical evolution, such as on the growth of instabilities and/or fragmentation adduced to these instabilities. In general, enhanced viscosity is likely to assist fragmentation and result in a number of small, fictitious fragments that could corrupt simulations and generate misleading results. \\

\begin{figure}
 \vspace*{5pt}
  \includegraphics[angle=270,width=9.cm]{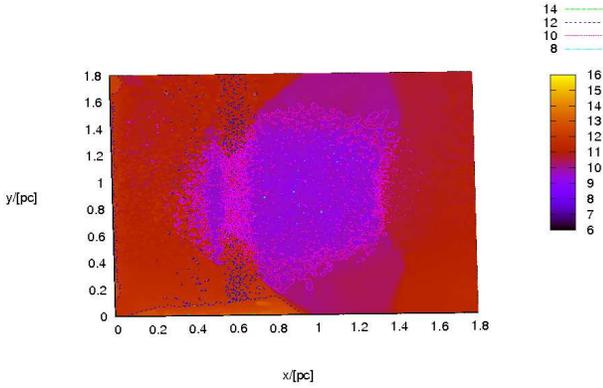}
 \caption{A plot showing the spatial distribution of vorticity, where the quantity  $(\nabla\times\nabla^{2}\textbf{v})$ has been plotted in logarithmic units over the $x-y$ plane; contours for the vorticity have been overlaid on the rendered image. It can be seen that shocked gas in general, shows greater dissipation of vorticity $(t\sim 0.017 \mathrm{Myrs})$.}
\end{figure}

\begin{figure}
 \vspace*{5pt}
  \includegraphics[angle=270,width=7.5cm]{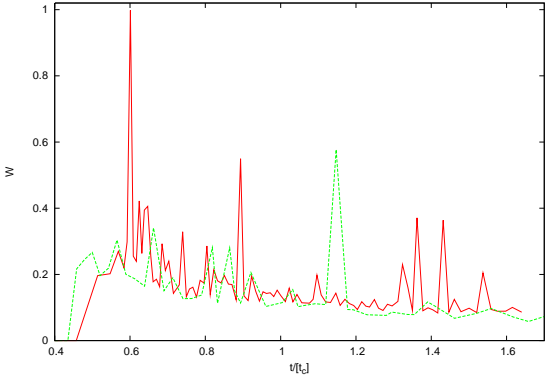}
 \caption{The time evolution of net vorticity in cases 1 (red curve) and 5 (green curve). Observe that a larger viscosity in case 5, evident from the spike at $t\sim 1.15 t_{cr} (\sim 7.5\times 10^{-3} \mathrm{Myrs})$, produces greater vorticity in the shocked cloud.}
\end{figure}

\subsection{Cases 6, 7 and 8}
\emph{Testing convergence: Resolution and artificial conductivity}\\
   These two cases are provisioned for additional discussion due to two crucial problems that could call into question the tenability of the preceding work. \\ (i) A pertinent issue in contemporary computational astrophysics is the ability of a numerical method to resolve hydrodynamical instabilities. Both {\small SPH} and the adaptive mesh refinement ({\small AMR}) codes have had considerable difficulty in resolving instabilities such as the KHI and the RTI (Springel 2010), constraining the discussion to {\small SPH} alone, problems such as segregation of regions with steep density contrast, or clumping of particles have been demonstrated (e.g. Agertz \emph{et al.} 2007; Junk \emph{et al.} 2010), and partially remedied (e.g. Price 2008; Wadsley \emph{et al.} 2008; Read \emph{et al.} 2010). While additional dissipative terms to aid fluid mixing, the artificial conductivity (AC), prescribed by Price (2008) has been, to some extent, successful in alleviating the problem related to segregation of fluid layers in purely hydrodynamical simulations, the KHI in this case has been demonstrated to grow on a longer timescale (e.g. Hubber \emph{et al.} 2010b); the switch, however, appears ineffective in presence of self-gravity as has been discussed above (\S 3.3, under cases 1 and 3). Figure 7 is a rendered density plot showing the shearing interaction between the slab and the surface of the cloud. The plot in the left and right-hand panels respectively correspond to the cases 1, and 7, described in Table 1 above; the former employs the AC prescription which is turned off in the latter. A visual examination alone readily demonstrates the similarity in the underlying structure.

(ii) Another crucial problem is that of particle clumping, the so called tensile instability, symptoms of which we had previously reported in {\small SPH} simulations of dissimilar clouds (see case 3 of Anathpindika 2010). Read \emph{et al.} 2010 have suggested an improved kernel along with a prescription for a large number of {\small SPH} neighbours, typically in excess of 400, to improve sampling of the kernel core and repel particles with vanishingly small separation. However, this prescription is likely to increase the smoothing length, at the expense of spatial resolution, as suggested by Eqn. (1) above. In the present enterprise we have therefore opted to retain the choice of 50 neighbours. Also, the physical conditions responsible for the tensile instability are unclear. Clumping of {\small SPH} particles is a rather localised occurrence, that will primarily manifest itself on the scale of a smoothing length, the structure reported in the simulations discussed here, however, is contiguous, spanning a few smoothing lengths. We therefore believe that the structure observed in the shocked cloud is physical, and corroborate our claim by repeating the simulation in case 1 by adopting a much larger number of particles; see Table 1 for physical details.

The plot in the top panel of Fig. 12 shows a rendered density plot of the collision sequence for case 6, the one with the highest resolution. The plots are coeval to those for case 1 shown in Fig. 5 above. It is clear that the overall morphology of the shocked cloud, though similar to cases discussed above, shows more filamentary structure and clumping in comparison to that in case 1.  Enhancement in structure formation is also visible in the density PDF plotted for this case in the bottom panel of Fig. 14. As expected, this PDF is broadly identical to that for case 1 shown in the top left-hand panel of Fig. 13, except in the density range $\sim(10^{3}-10^{4.5})$ cm$^{-3}$. The PDF for case 1 in this interval is stunted whereas that in the present case has a more pronounced peak, which again points to the promiscuous formation of dense structure in this case. We adduce this to an increase in spatial resolution which in this case is at least by a factor of two as described previously in \S 2.
 While on the one hand the interface between the cloud and the impinging slab is KH unstable, the Jeans instability is crucial to structure-formation, and eventually, to the formation of prestellar cores in this structure. Resolution of this instability is therefore our concern. As reported by several authors (e.g. Bate \& Burkert 1997; Hubber \emph{et al.} 2006) in the past, insufficient resolution tends to damp the growth of perturbations, which in the problem considered here, are purely gravo-thermal in nature. The ratio of the Jeans length, $\lambda_{J}$, to $h_{avg}$ in the relatively lower resolution cases, $\mathcal{R}\equiv \lambda_{J}/h_{avg}\sim 6$, is conservative in view of the Truelove criterion for resolving the Jeans instability; $\mathcal{R}\gtrsim 8$ to satisfy the Truelove criterion. Having increased the number of particles within the cloud by a factor of eight (case 6), $\mathcal{R}\equiv \lambda_{J}/h_{avg}\sim 13$, so that the Jeans length could possibly be resolved.   \\ \\ 
\textbf{\emph{\textbf{Case 8 - Lowest resolution}}}  \\
The number of particles within the cloud for this case are reduced by a factor of two, so that the average smoothing length, $h_{avg}$, defined in \S 2 above is $\sim$1.25 times that in case 1 so that $\mathcal{R}\equiv\frac{\lambda_{J}}{h_{avg}}\sim  4$, and therefore, obviously violates the Truelove criterion to resolve Jeans instability. The shocked cloud lacks sufficient spatial resolution and consequently shows hardly any signs of fragmentation which is not surprising at all. The bottom panel in fig. 12  shows a cross-sectional plot of the shocked cloud in this case.

These occurrences at the two extreme choices of $N_{cld}$, adumbrated respectively, by cases 6, and 8, suggest that poor spatial resolution tends to suppress fragmentation so that the number of particles in the computational domain should be increased as much possible. The point has been emphasised by numerous authors in the past (e.g. Agertz \emph {et al.} 2007; Commer\c{c}on \emph{et al.} 2008, and several references therein). An associated problem is of the stability of the preshock cloud against perturbations.
The cross-sectional plots of the shocked cloud shown in Figs. 5, 9, and the top panel of 12, for instance, show that density structure appears at the rear end of the cloud even before that part is engulfed by the impinging shock. We therefore conclude that instabilities observed in the shocked cloud are triggered purely by white noise, and it appears, one may have to use at least a billion particles to possibly ensure stability. We discuss the stability of the density field in Appendix A below where a conservative lower-limit on $N_{cld}$ has also been derived. 

\begin{figure*}
\vbox to 100mm{\vfil 
  \includegraphics[angle=270,width=18.cm]{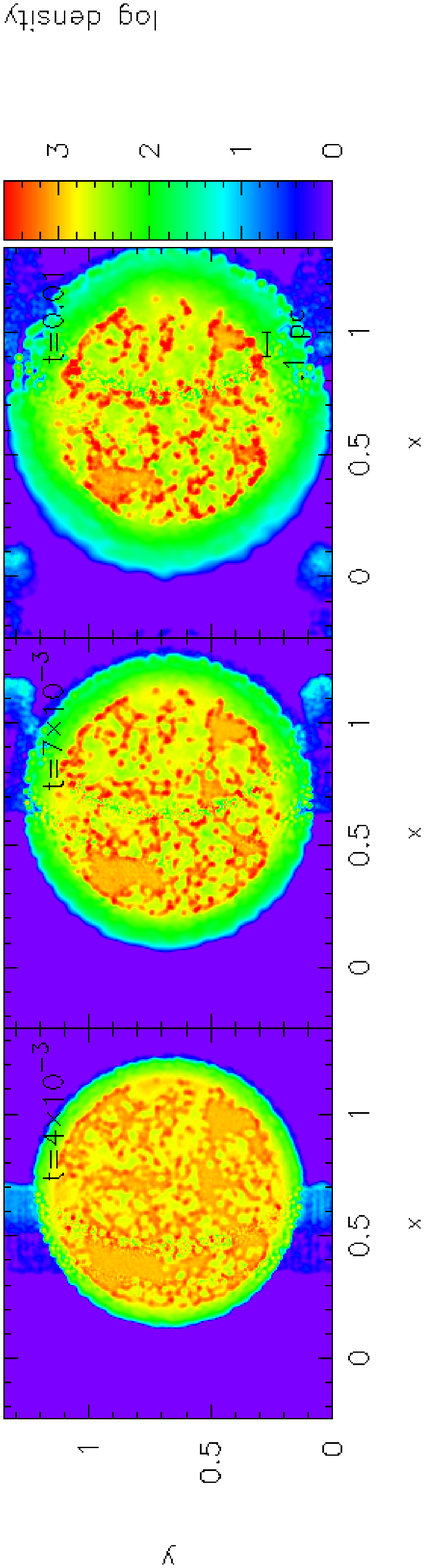}
  \includegraphics[angle=270,width=10.cm]{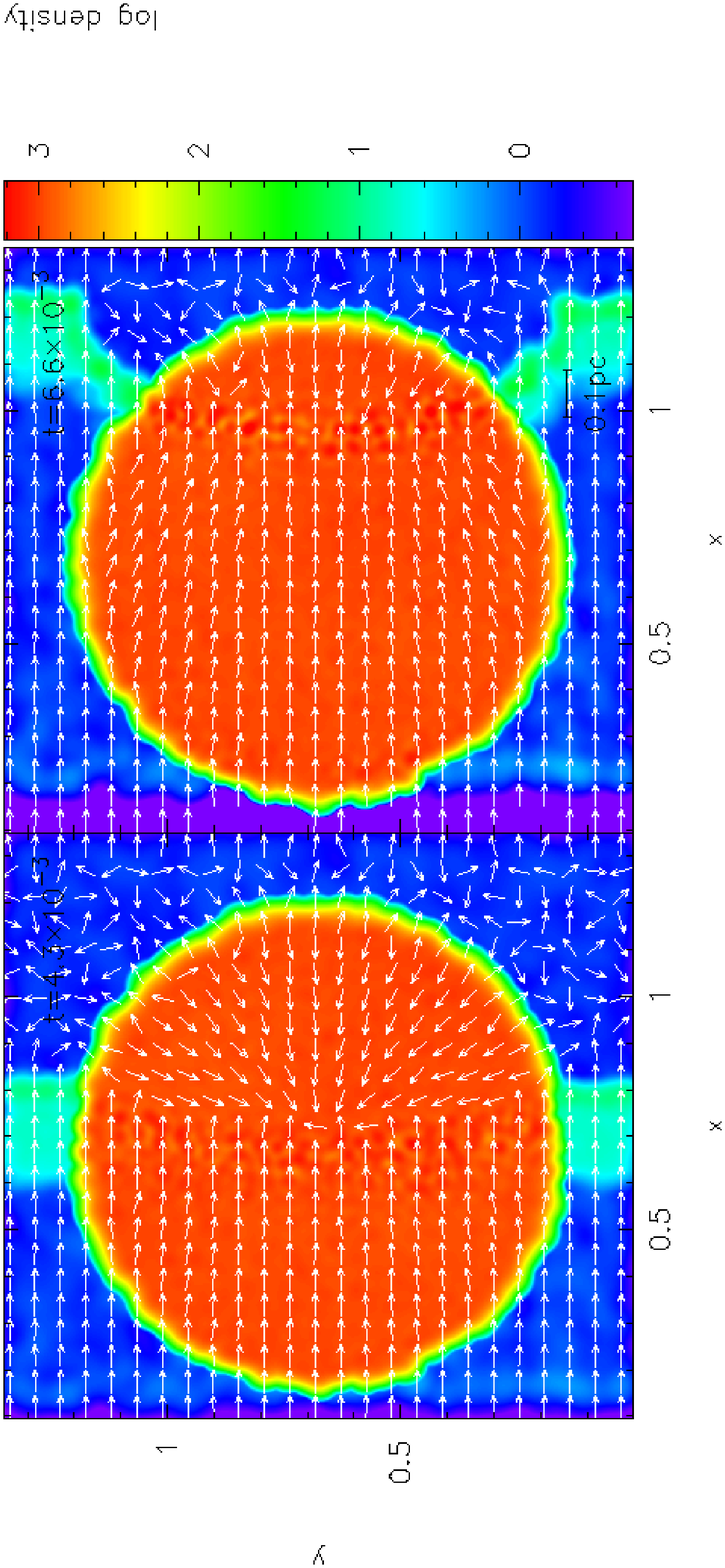} 
\caption{\emph{Top panel} A montage showing cross-sectional rendered images of the shocked cloud from the best resolved simulation ($N_{tot}=1.2\times 10^{6}$), case 6 in Table 1. The plot is coeval relative to that shown in Fig. 5 above, for case 1. See text for further description. \emph{Bottom panel} A montage showing a density cross-section of the shocked cloud in case 8. This is the case with lowest resolution and shows hardly any evidence of fragmentation ($N_{tot} = 80,000$).} \vfil} 
\end{figure*}

\subsection{Density PDFs and the core mass function}

The PDF of a variable provides information about its distribution in sub-intervals spanning the whole range. The PDF in each of the five cases discussed above are distinctly bimodal which indicates that part of the underlying gas is distributed in clumps and filaments, denser than the rest of it; see figs. 13 and 14. The remainder of the gas is unbound and therefore rarefied. The volume filling factor, $f$, of the dense structure within the shocked cloud can be roughly estimated by taking the ratio of the observed peak in the density PDF with the density of the precollision cloud. This yields $f\sim$10 \%, in other words 90 \% of the shocked gas ends up in rarefied holes, or voids, which is in agreement with the value reported by \citet{b12} for fractal clouds. This observed segregation of the dense and the rarefied medium is an important characteristic of the bistable model of the interstellar medium. While this is true for the denser precollision clouds in cases 1, 3, and 6, the rarefied clouds in cases 2, and 4, after being shocked, show some what contradictory behaviour. The PDF in case 2 is peaked at a density close to the density of the precollision cloud, which suggests, the isothermal shock has had little effect on the distribution of precollision density within the cloud. On the other hand, the adiabatic shock in case 4 leads to a more significant density enhancement in the shocked cloud. The shocked gas in this latter case is warm, at roughly a few hundred Kelvin which is more than an order of magnitude larger than the gas temperature in case 2. The pressure-distribution  within the shocked cloud plotted in the bottom right-hand panel Fig. 8 shows that the cloud in this case has shrunk by about 10\% which is also responsible for density enhancement within the cloud that results in the PDF plotted in the bottom right-hand panel of Fig. 4. While the incident shock raises the pressure within the cloud, it drops off rapidly near edges.

Fragmentation in warm gas is likely to be delayed, leading to greater contiguity of structure. We note that, the PDF in these test cases, at the high density end, shows an approximate lognormal nature. Our work therefore reinforces the importance of interstellar shocks in the process of stellar birth as has been pointed out by numerous workers in the past (e.g.V{\`a}zquez-Semadeni (1994); Padoan \& Nordlund (2002); V{\`a}zquez-Semadeni \emph{et al.} (2008); Hennebelle \& Chabrier (2008) ). We need not be unduly concerned about the tail of the PDFs extending into spectacularly low densities, as it is likely to be part of the warm ICM. At this point it would therefore be useful to briefly consider the spectrum for the two phases. The energy density of the turbulent field in Fourier space, $E(k)$, and that in the real space is related as,
\begin{displaymath}
\int_{0}^{\infty} E(k)dk = \frac{1}{2}\sum_{i} v_{i}^{2},
\end{displaymath} 
$k$ being the wavenumber. The summation on the right-hand side above runs over all {\small SPH} particles. The smallest spatial scale on which energy could possibly be dissipated is the {\small SPH} smoothing length, $h_{i}$, so that $k_{i}\sim 1/h_{i}$. The spectrum so calculated is plotted in Fig. 15 which also shows a segregation of the two gas phases. While the rarefied gas produces a Kolmogorov like spectrum, $\propto k^{-1.7}$, that for the dense gas is much steeper which suggests that the Kolmogorov approximation of inviscid fluids breaks down in high density regions.

We close this discussion on PDFs by briefly considering the effect of higher viscosity in case 5, on the distribution of post-shock density. As in the previous cases, this PDF is also bimodal, but peaked at the lower end of the density distribution. This again suggests that there is greater contiguity of structure, that may fragment only after becoming sufficiently massive. This PDF, as it evolves in time will gradually move towards higher densities, however, an important constraint here, and particularly, from the perspective of star-formation, is the timescale over which this shift may be effected. The PDF shown in the top panel of Fig. 14 is plotted at a similar epoch as in the previous cases, and it appears that a larger viscosity tends to slow down the process of turbulent fragmentation. This leads us to two possible scenarios : (a) Turbulence within gas tends to produce substructure on a rather short timescale, and (b) quiescent gas on the other hand, tends to agglomerate more gas and become sufficiently massive before it could possibly become Jeans unstable. The post-collision shock in case 5 becomes more or less stationary due to higher viscous dissipation, and so gas within the cloud remains by and large quiescent, qualifying for category (b) above. The proposition under (b) is less likely to explain the formation of stars for it will tend to delay the onset of the star-formation, returning a timescale grossly inconsistent with that commonly reported by surveys of young star-forming regions. 

\begin{figure*}
\vbox to 110mm{\vfil 
  \includegraphics[angle=270,width=13.cm]{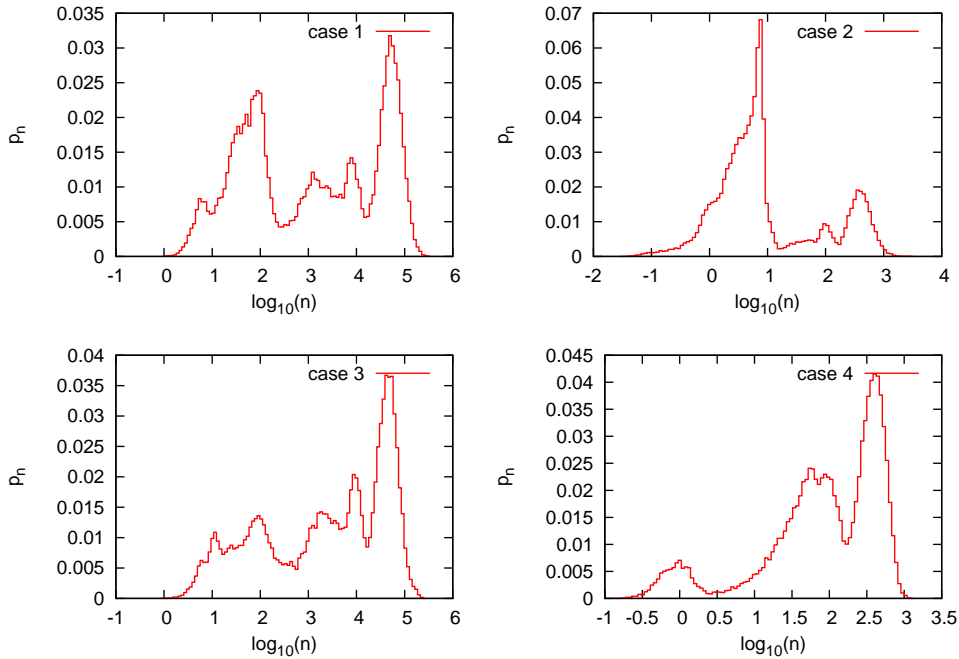}   
\caption{The density probability distributions for the first 4 cases. The bimodal nature of PDFs shows that the propagation of a shock leads to segregation of dense and rarefied gas phase. \emph{See text for description.}}\vfil} 
\end{figure*}

\section{Conclusions}

Starting with a relatively stable MC, we have demonstrated that turbulence injected due to the impact of a thin shell produces structure within the post-collision cloud. And crucially, this fragmentation occurs on a rather short timescale, comparable to the local crossing time, $t_{c}$, defined by Eqn. (3) above. Interestingly, the filling factor, $f$, for dense structure is consistent with that reported for fractal clouds in the interstellar medium, which reinforces the idea that interstellar shocks could be vital in producing the observed dense structure in the galactic disk; prestellar cores, under appropriate physical conditions, could condense in this structure. The degree of fragmentation, however, depends on thermodynamic details of the process, and in general warm, and/or rarefied gas, as expected, is observed to be quiescent. The results of our work are particularly interesting in view of the recently reported sub-mm observations of the Gould belt with the Herschel space observatory that reveals a number of prestellar cores in several dense, gas filaments (e.g. Andre et al. 2010). We have tested the convergence of our simulations by repeating the test with more than a million particles, where filaments in the shocked cloud are even more promiscuous; see top panel of Fig. 12; insufficient resolution, however, suppresses structure formation as can be seen in the bottom panel of Fig. 12. Starting with a relatively stable molecular cloud, it has been demonstrated that runaway of growth of perturbations in the density field of the cloud lead to formation of small clumps, and more contiguous filaments. Any density fluctuations arise purely from white noise.

Turbulence in the physical world is complex, and  unlikely to be adequately represented by a single power-spectrum, such as the Kolmogorov spectrum, and more crucially, the Kolmogorov analysis was prescribed for inviscid fluids whereas real fluids are viscous. Models such as the one tested here provides naturally a more realistic turbulent field with vortices which permits a study of the formation of dense structure from stable initial conditions. Vortices appear to play a key role in viscous dissipation of energy, evident from the plots in Figs. 11 and 12. The montage in the top panel of Fig. 12 in fact demonstrates that substantial energy dissipation, manifested by reduced vorticity, precludes structure formation, the boundedness of which, of course, depends on the extent of dissipation. We might thus, be in a better position to answer crucial questions related to the rate and efficiency of star-formation, all of which depend directly on the ambient conditions in the star-forming regions. The post-collision shock readily generates a power-law distribution of interstellar gas as has also been demonstrated by numerous authors in the past, while in the cases discussed above, segregation between the dense and rarefied phases of the gas is also evident from the PDFs plotted in Figs. 13 and lower panel of Fig. 14. This is further corroborated by the energy power spectrum for the two phases plotted in Fig. 15. The spectrum for the diffuse gas is Kolmogorov like, while that for the dense gas is much steeper which agrees with our observation in an earlier related work (Anathpindika 2010).

The model tested here, based on the well-known concept of turbulent fragmentation, suggests that the process leading to the assembly of a prestellar core is rapid, and may not last longer than a crossing time. The filaments and clumps produced in the present work could have been allowed to evolve further. But for the limited computational resources at our disposal, any such ideas will have to be deferred for future exploration. As a divestment to the central problem, we have also attempted to study the influence of numerical viscosity on the post-shock dynamics. We conclude that too much viscosity, and in fact the standard {\small SPH} viscosity, is likely to produce misleading results by inducing spurious fragmentation.

\begin{figure}
 \vspace*{4pt}
  \includegraphics[angle=270,width=7.5cm]{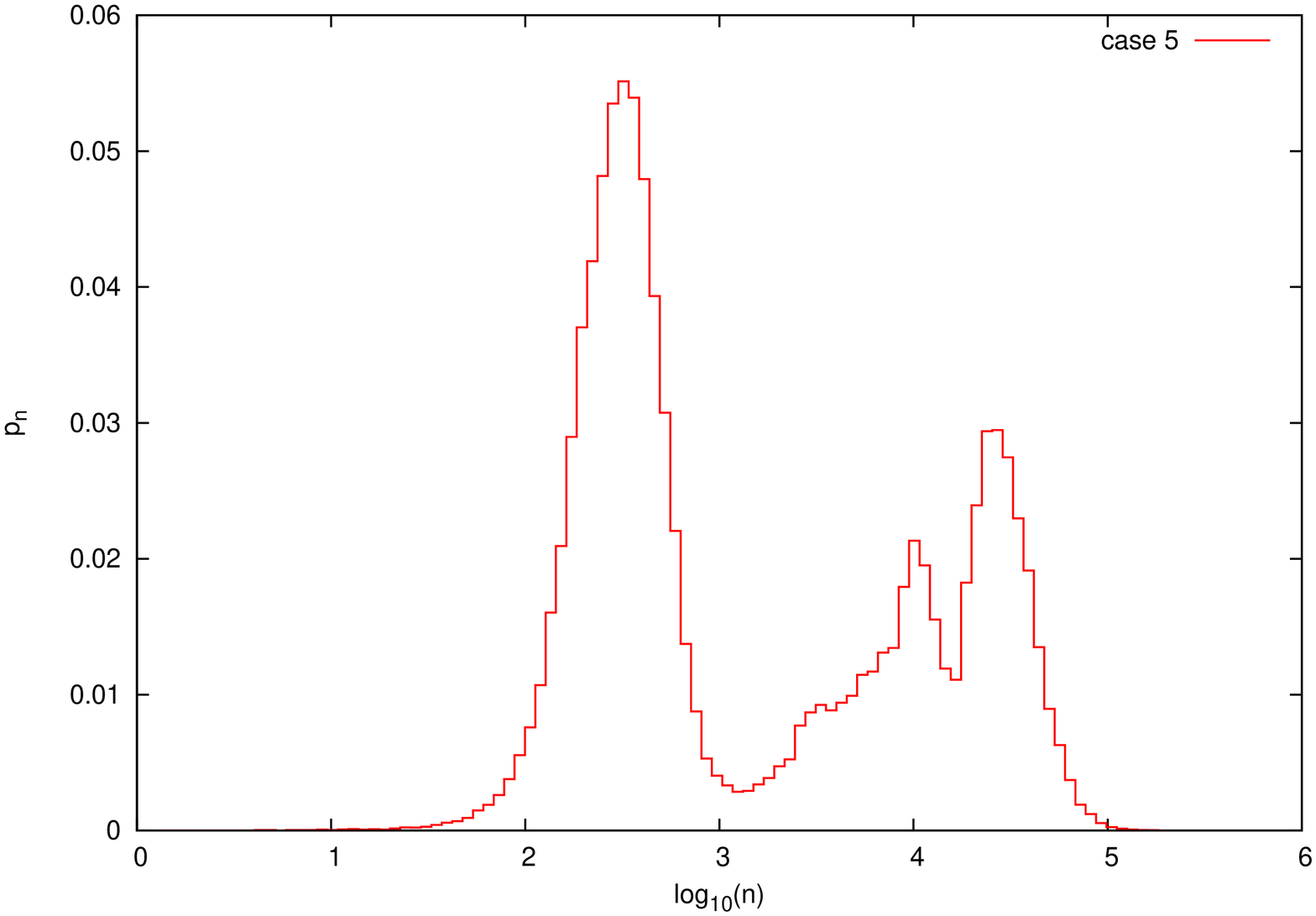}
  \includegraphics[angle=270,width=7.5cm]{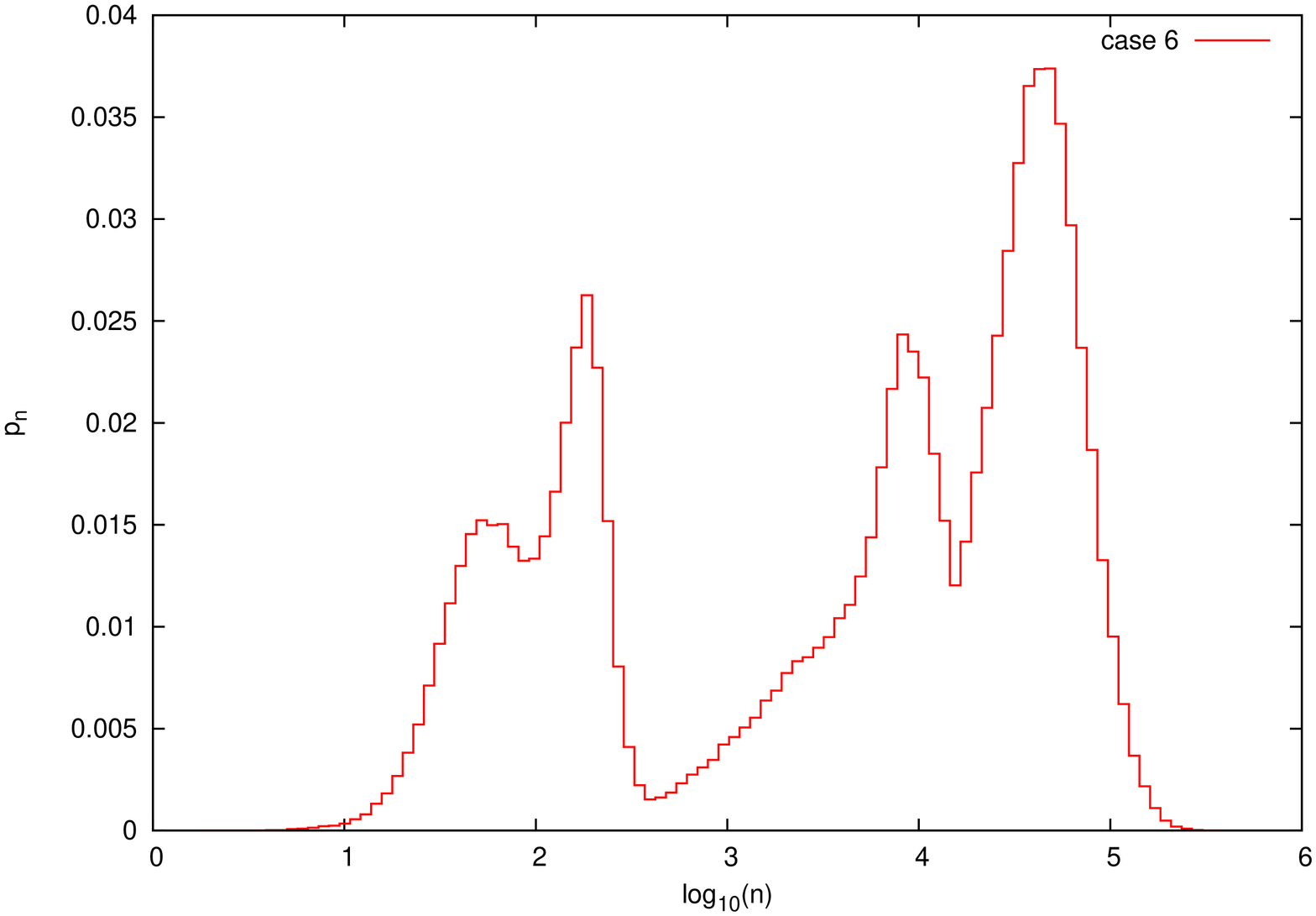}
 \caption{\emph{Top panel :} The density PDF for case 5. This PDF, as for the first 4 cases, is also bimodal but peaks at a much lower density although the epoch of this plot is the same for all cases. \emph{Bottom panel :} The density PDF for case 6, which like other cases also shows a bimodal distribution, but is more pronounced for intermediate densities in comparison to that for case 1 plotted in Fig. 13 which demonstrates preferential distribution of matter in contiguous structure. }
\end{figure} 

\begin{figure}
 \vspace*{4pt}
  \includegraphics[angle=270,width=7.5cm]{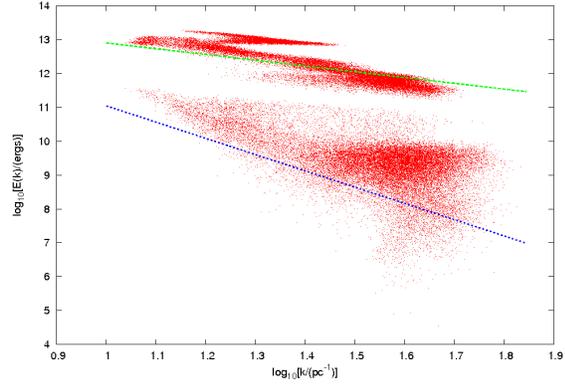}
 \caption{The power spectrum for the two gas phases. Respective power-laws have also been plotted for comparison; the power-law for diffuse gas shown by a green line, $\propto k^{-1.7}$, is in agreement with the Kolmogorov spectrum, while that for dense gas, is much steeper shown by a purple line, $\propto k^{-4.5}$.}
\end{figure} 

\section*{Acknowledgments}
We wish to thank an anonymous referee for insightful comments that helped us emphasise a few vital points, particularly in connection with structure formation in shocked gas. Special thanks to David Hubber and Chris Batty for making available a fresh copy of the {\small SPH} code, Seren. Simulations discussed in this paper were partially performed using the supercomputing cluster maintained by the C-DAC at Bangalore, India. Dr Anathpindika is supported by a post-doctoral fellowship at the Indian Institute of Astrophysics, Bangalore. This project was partially conceived during the doctoral research of the author at the School of Astronomy and Physics, Cardiff university.

\appendix
\section{Stability analysis of the {\small SPH} density field}
Let us begin with the equation for {\small SPH} density of the $i^{th}$ particle, $\rho_{i}(r_{i})$, having position vector $r_{i}$
\begin{equation}
\rho_{i}\equiv\rho(r_{i}) = m\sum_{j} W_{ij}(s,h_{i}),
\end{equation}
where $s=\vert\textbf{r}_{i} - \textbf{r}_{j}\vert/h_{i}$, $W(s,h_{i})$ is the smoothing kernel and the summation is over number of all {\small SPH} neighbours, $N_{neibs}$. Individual particles are assumed to have identical masses, $m$. The kernel employed in the simulations discussed above is
\begin{equation}
W(r,h) = \frac{21}{256\pi h^{3}}(2-s)^{2}(4-s^{2}); s < 2
\end{equation}
Thomas \& Couchman (1992).
 The time-derivative of the {\small SPH} density, $\rho$, is simply
\begin{equation}
\frac{d\rho_{i}}{dt} = m\sum_{j} \frac{dW_{ij}(s,h)}{dt},
\end{equation}
and 
\begin{displaymath}
\frac{dW_{ij}(s,h)}{dt} = \frac{dW_{ij}}{ds}\frac{\partial s}{\partial t} = \frac{v_{ij}}{h_{i}}\frac{dW_{ij}}{ds},
\end{displaymath}
which on substituting in Eqn. (A-3) we get
\begin{equation}
\frac{d\rho_{i}}{dt} = \frac{m}{h_{i}}\sum_{j} v_{ij}\frac{dW_{ij}}{ds},
\end{equation}
which is simply the equation of continuity, and $v_{ij}$ is the signal velocity between the particle pair $(i,j)$. Our interest lies in exploring the possibility of structure formation due to a small perturbation in the {\small SPH} density field so we adopt the standard procedure where the perturbation is assumed to be oscillatory with angular frequency $\omega$, 
\begin{equation}
\rho_{i} = \rho_{i}^{0}e^{-i\omega t}.
\end{equation} 
$\rho_{i}^{0} = \frac{21m}{256\pi h^{3}_{i}}$ is the true density of the {\small SPH} particle ($s=0$ in Eqn. (A-2)). Making necessary substitutions, Eqn. (A-4) becomes
\begin{displaymath}
-i\omega \rho_{i}^{0}(\cos(\omega t) -i\sin(\omega t)) = \frac{21}{64\pi h^{4}_{i}}m\sum_{j} v_{ij}(3s^{2} - 4),
\end{displaymath}
and comparing real parts on either sides we wind up with,
\begin{displaymath}
-\omega\sin(\omega t)= 4v_{ij}(3s^{2} - 4)
\end{displaymath}
which for an infinitesimally small perturbation becomes
\begin{displaymath}
-\omega^{2}t\sim 4v_{ij}(3s^{2} - 4)
\end{displaymath}
or
\begin{equation}
\omega \sim 2\Big[\frac{v_{ij}(4-3s^{2})}{t}\Big]^{1/2}.
\end{equation}
This expression leads us to the following three cases : \\ \\
\emph{\textbf{Case 1 }} $\mathbf{\omega^{2} > 0}$; there exists a real root so that a density perturbation defined by Eqn. (A5) will remain oscillatory. \\ \\
\emph{\textbf{Case 2}} $\omega^{2} < 0$; the root is imaginary which suggests that any perturbation, according to Eqn. (A5), will grow exponentially in time.\\ \\
\emph{\textbf{Case 3}} $\omega^{2} = 0$; the condition when the density field is likely to remain steady. \\ \\
Equation (A6) may be used to derive a lower limit on the maximum number of particles, $N_{cld}$, in a cloud. The condition to avoid growth of density perturbations defined by case 3 above implies $s=2h^{1/2}$, or $h = (0.5r_{ij})^{2/3}$ but $r_{ij}\propto (N_{cld})^{-1/3}$ so that $h\propto (N_{cld})^{-2/9}$. Even by the most conservative estimates, $N_{cld}\sim 10^{9}$ before the average smoothing length is reduced by about 29\% to satisfy the equality. Evidently, any possible gain in stability will exact enormous computational expense. Thus, for all practical purposes the initial set of {\small SPH} particles, as qualified by \emph{case 1} above, is likely to be susceptible to numerical instability.

\bsp
\label{lastpage}

\end{document}